# Quasi-one-dimensional superconductivity in the pressurized charge-density-wave conductor HfTe$_3$

Z. Y. Liu[1,2,#], J. Li[2,3,#], J. F. Zhang[4,#], J. Li[2,3], P. T. Yang[2], S. Zhang[2,3,*], G. F. Chen[2,3], Y. Uwatoko[5], H. X. Yang,[2,3] Y. Sui[1,*], K. Liu[4,*], and J.-G. Cheng[2,3,*]

[1]*School of Physics, Harbin Institute of Technology, Harbin 150001, China*

[2]*Beijing National Laboratory for Condensed Matter Physics and Institute of Physics, Chinese Academy of Sciences, Beijing 100190, China*

[3]*School of Physical Sciences, University of Chinese Academy of Sciences, Beijing 100190, China*

[4]*Department of Physics, Beijing Key Laboratory of Opto-electronic Functional Materials & Micro-nano Devices, Renmin University of China, Beijing 100872, China*

[5]*Institute for Solid State Physics, University of Tokyo, Kashiwa, Chiba 277-8581, Japan*

\# These authors contributed equally to this work.

*E-mails: jgcheng@iphy.ac.cn; szhang@iphy.ac.cn; suiyu@hit.edu.cn; kliu@ruc.edu.cn

## Abstract

HfTe$_3$ single crystal undergoes a charge-density-wave (CDW) transition at $T_{CDW}$ = 93 K without the appearance of superconductivity (SC) down to 50 mK at ambient pressure. Here, we determined its CDW vector ***q*** = 0.91(1) *a*\* + 0.27(1) *c*\* via low-temperature transimission electron microscope and then performed comprehensive high-pressure transport measurements along three major crystallographic axes. Our results indicate that the superconducting pairing starts to occur within the quasi-one-dimensional (Q1D) -Te2-Te3- chain at 4-5 K but the phase coherence between the superconducting chains cannot be realized along either the *b*- or *c*-axis down to at least 1.4 K, giving rise to an extremely anisotropic SC rarely seen in real materials. We have discussed the prominent Q1D SC in pressurized HfTe$_3$ in terms of the anisotropic Fermi surfaces arising from the unidirectional Te-5p$_x$ electronic states and the local pairs formed along the -Te2-Te3- chains based on the first-principles electronic structure calculations.

**Keywords:** HfTe$_3$, charge density wave, superconductivity, pressure effect

## Introduction

Charge density wave (CDW) and conventional superconductivity (SC) are both caused by strong electron-phonon coupling and Fermi surface (FS) instabilities, and thus the interplay between these two collective electronic phenomena has been a subject of extensive investigations over the past decades [1]. The former usually occurs in the low-dimensional materials and involves a periodic modulation of the conduction electron density and the underlying lattice; this will open a gap over part of the FS and thus lead to reduced conductivity. In contrast, bulk SC is a macroscopic quantum phenomenon that requires phase coherence in three dimensions and it features a FS gap that results in infinite conductivity. As such, these two electronic orders are usually competing with each other. It is frequently observed in the low-dimensional CDW conductors that suppression of CDW order leads to a bulk SC. Here, we report on a rare case that an intrinsic quasi-one-dimensional (Q1D) filamentary SC emerges and persists in a relatively wide temperature/pressure range when the CDW order of HfTe$_3$ along the same crystallographic direction is completely suppressed by pressure.

As shown in Fig. 1(a), HfTe$_3$ crystallizes in the TiS$_3$-type structure with monoclinic space group $P2_1/m$, featured by two kinds of chain-like units, *i.e.* the infinite prismatic (HfTe$_3$)$_\infty$ chains along the *b*-axis and the almost linear chains of -Te2-Te3- with alternating short (2.80 Å) and long (3.07 Å) distances along the *a*-axis [2]. The two-dimensional (2D) prism slabs in the *ab* plane are stacked along the *c*-axis through weak van der Waals (vdW) interactions. These quasi-1D and 2D structural characters have a profound impact on the physical properties of HfTe$_3$ at ambient pressure. Resistivity measurements on high-quality HfTe$_3$ single crystals reveal pronounced anisotropic behaviors and obvious anomalies in both $\rho_a(T)$ and $\rho_c(T)$ but a very weak kink-like anomaly in $\rho_b(T)$ at around 93 K [2]. In analog with the sister compound ZrTe$_3$ [3-19], the transport anomaly at $T_{CDW}$ = 93 K has been ascribed to the formation of a CDW order within the *ac*-plane [6]. Since the $\rho_c$ is about two orders of magnitude higher than $\rho_a$, the CDW should have a dominate 1D character along the -Te2-Te3- infinite chains. However, the wavevector of the CDW order has not been determined experimentally.

For high-quality single crystalline HfTe$_3$ samples, no SC was detected down to 50 mK at ambient pressure [2]. In contract, the polycrystalline HfTe$_3$ sample shows the coexistence of the CDW order below $T_{CDW} \approx 80$ K and filamentary SC with $T_c \approx 2$ K at ambient pressure [2]. The resistivity anomaly around $T_{CDW}$ is much reduced in comparison with that of single crystal, implying that the emergence of SC should be correlated with the suppression of CDW, presumably due to the presence of chemical disorders or defects in the polycrystalline samples. However, the nature of the filamentary SC without magnetic and thermodynamic signatures and its relationship with CDW remains elusive so far. For the polycrystalline sample, Denholme *et al*. [20] have performed preliminary high-pressure (HP) resistivity measurements up to ~1 GPa; they observed pressure-induced enhancement of the CDW transition, but the evolution of SC and its relationship with respect to CDW was not reported [20]. Considering the strong anisotropy of the physical properties associated with the Q1D nature of the crystal structure, it is imperative to carry out a comprehensive HP study on single-crystalline samples in order to unveil the nature of the filamentary SC and its interplay with CDW.

In this work, we first determined experimentally the CDW vector $\boldsymbol{q}$ = 0.91(1) $a^*$ + 0.27(1) $c^*$ of HfTe$_3$ by low-temperature transmission electron microscope (TEM) at ambient pressure, and then performed a comprehensive HP transport study on the HfTe$_3$ single crystals along three major crystallographic axes. Intriguingly, we observed the emergence of an intrinsic filamentary SC in $R_a(T)$ at $T \leq$ 4-5 K when the CDW of HfTe$_3$ is suppressed completely by $P \geq$ 5 GPa, whereas no clear sign of SC is observed down to 1.4 K in $R_b(T)$ and $R_c(T)$, demonstrating an extraordinary case of Q1D SC rarely seen in real materials. We have analyzed the excess conductivity according to the theory of Aslamazov and Larkin (A-L) [21] and compared our results with those of ZrTe$_3$ [3-19,22]. Finally, we discussed the observed Q1D SC in terms of the anisotropic FSs and the local pairs formed along the -Te2-Te3- chains parallel to the $a$-axis based on the first-principles electronic structure calculations.

## Results

### A. Characterizations at ambient pressure

The single-crystal XRD pattern of HfTe$_3$ shown in Fig. 1(b) displays only the sharp (00$l$) diffraction peaks, which confirm that the single crystals are grown along the $ab$ plane and stacked along the $c$-axis. The $c$-axis lattice constant is determined to be $c$ = 10.0519 Å, in consistent with the previous report [2]. As shown in the inset of Fig. 1(b), the full width at half maximum (FWHM) of the (006) Bragg peak is about 0.2°, implying a high quality of the crystals. Within the plate-like crystal surface, the $a$ and $b$ axes are determined to be perpendicular and parallel to the crystal edges, respectively, by the Laue back diffraction. As shown in the inset of Fig. 1(c), the HfTe$_3$ single crystal shows fringes extending along the $b$-axis, which is parallel to the crystal edge. Thus, the $a$- and $b$-axis that are orthogonal with each other within the $ab$ plane can be distinguished visually for the as-grown crystals. For this sample, we measured the temperature dependence of anisotropic resistance $R(T)$ along the $a$- and $b$-axis, respectively. As seen in Fig. 1(c), the $a$-axis resistance exhibits a pronounced hump anomaly at $T_{CDW}$ = 93 K associated with the CDW transition, while the $b$-axis resistance shows a very weak kink-like anomaly at $T_{CDW}$ without any trace of superconducting transition down to 1.8 K. These results agree well with the previous report [2]. The residual resistivity ratio, RRR ≡ $R$(300 K)/$R$(2 K) along the $b$-axis is calculated to be 62, which also highlights the high quality of the studied single-crystal sample. We have thus selected samples with similar RRR values for the HP measurements as shown in the Supplementary Fig. 1.

Although it is expected that HfTe$_3$ should form a similar CDW as that of ZrTe$_3$ with $\boldsymbol{q}$ = 0.93 $a^*$ + 0.33 $c^*$, the $q$-vector of HfTe$_3$ has not been determined experimentally so far. Here we employed a low-temperature TEM to directly determine the CDW vector of HfTe$_3$ single crystal. Figure 2(a) and (b) show the selected area electron diffraction

(SAED) patterns of HfTe$_3$ taken along the [001] zone axis at room temperature and 26 K, respectively. The pristine lattice at room temperature has a monoclinic structure with space group $P2_1/m$, and the satellite spots can only be identified below the CDW phase transition temperature. As shown in the inset of Fig. 2(b), the intensity of satellite spot goes up as the scattering angle increases and reaches a maximum at 17.5 mrad, indicating that the structural modulation vector $q$ has both $a^*$ and $c^*$ components, i.e., $q = q_a a^* + q_c c^*$. The extra Laue zones (indicated by blue arrows) presented in the [001] CBED pattern, Fig. 2(c), further confirm that the modulation vector does not lie in the $ab$-plane.

In order to determine the value of $q_a$ and $q_c$, the intensity distributions of CDW spots/rings in Fig. 2(b) and (c) are summarized in Fig. 2(d) and (e), respectively. Figure 2(d) shows the line intensity profile from spot $(\bar{6}00)$ to $(\bar{10}00)$, in which the intervals between adjacent Bragg spots ($L_{Bragg}$) and the distance from one Bragg spot to the next-nearest CDW neighbor ($L_{CDW}$) are labeled. Using $q_a = L_{CDW}/L_{Bragg}$, $q_a$ is calculated to be 0.91(1). On the other hand, the value of $q_c$ can be estimated with the aid of the geometric configuration presented in Fig. 2(e). The positions of the rings in Fig. 2(c) are directly related to the interception of the Ewald sphere with the Laue zone layers. For high-energy electrons, we can write an expression for the angle $\theta$ in Fig. 2(e): $\sin\theta = \frac{\Delta K/2}{r} = \frac{h}{\Delta K}$. Using the equation $(\Delta K)^2 = h^2 + L^2$, we have $h = r - \sqrt{r^2 - L^2}$, where $r$ is the radius of the Ewald sphere ($r$ = 398.74 nm$^{-1}$ when the electron energy is 200 keV). The $L$ values for different rings can be obtained from the radial profile plotted in the low panel of Fig. 2(e). Then, by calculating $h$ for different Laue zones and the ratios (for example $q_c = h_{CDW}/h_{FOLZ}$, where $h_{CDW}$ and $h_{FOLZ}$ are the $h$ values for the innermost ring and the FOLZ ring respectively), $q_c$ is estimated to be 0.27(1). Thus, we can write $q = 0.91(1) a^* + 0.27(1) c^*$.

To further understand the temperature-dependent behavior of the structural modulation, we acquired SAED patterns at several different temperatures, as shown in Supplementary Fig. 2. It is evident that the superlattice spots (indicated by blue arrows) become dim at 81 K and totally disappear at 88 K, showing good agreement with the CDW phase transition temperature determined from the resistivity measurements. The result suggests that the superlattice spots observed in diffraction patterns are directly related to the formation of CDW phase.

It is worth noting that the satellite spots appeared in the electron diffraction patterns of ZrTe$_3$ at room temperature. Eaglesham et al. [6] believed that the satellite spots at room temperature indicated the composition modulation of ZrTe$_3$, which were caused by off-stoichiometry. This is also considered to be a strong evidence that the quality of ZrTe$_3$ crystals is difficult to control. For HfTe$_3$ single crystal, we can see that the SAED patterns do not show any satellite spots at room temperature, which indicates that the crystal structure of HfTe$_3$ single crystal is perfect and closer to the stoichiometry.

### B. Effect of pressure on the $a$-axis resistance

Fig. 3(a) shows the temperature dependence of the $a$-axis resistance $R_a(T)$ of the HfTe$_3$ single crystal (#1) under various pressures up to 2.0 GPa measured with a PCC. Details about the experimental setup for anisotropic resistance measurements can found in the Supplementary Fig. 4. At ambient pressure, $R_a(T)$ displays a hump anomaly due to the CDW formation at $T_{CDW}$ = 93 K, which can be determined from the minimum of d$R_a$/d$T$ as shown in the inset of Fig. 3(a). With increasing pressure, $T_{CDW}$ first moves to higher temperatures, reaching the maximum of ~128 K at 1.5 GPa, and then decreases gradually to 122.4 K at 2.01 GPa. The jump of $R_a(T)$ around $T_{CDW}$ is reduced gradually accompanying the suppression of CDW transition. Such a non-monotonic variation of $T_{CDW}(P)$ has also been observed in ZrTe$_3$,[12] and thus should have an origin in common. The initial enhancement of $T_{CDW}$ at low pressures indicates that the lattice compression first prompts the CDW formation, presumably due to the enhancement of electron-lattice coupling or the optimization of the nesting condition for the parallel FSs along the $a^*$-axis. Nonetheless, the CDW should be eventually suppressed by pressure as observed in ZrTe$_3$ [12].

The evolution of the CDW transition at $P$ > 2 GPa is further investigated by measuring $R_a(T)$ of a second HfTe$_3$ crystal (#2) with a CAC. The obtained $R_a(T)$ and the selected d$R_a$/d$T$ curves are displayed in Fig. 3(b). The $R_a(T)$ at 2.4 GPa is similar to that measured in PCC, and the $T_{CDW}$ has been further reduced to 93.8 K. When the pressure is increased to 4 GPa, the $T_{CDW}$ is quickly lowered to 43.3 K and the resistance anomaly at $T_{CDW}$ is significantly suppressed, as shown in Fig. 3(b). At 5.4 GPa and above, no resistance anomaly can be discerned any more at $T$ > 5 K, implying the complete suppression of the CDW transition in HfTe$_3$. In this pressure range, the magnitude of $R_a(T)$ at high temperatures decreases gradually with increasing pressure while that at low temperatures keeps almost constant.

Interestingly, we observed a gradual resistance drop at low temperatures for $P$ > 5.4 GPa, signaling the possible emergence of SC accompanying the elimination of CDW in HfTe$_3$. As shown in Fig. 3(c), the enlarged view of $R_a(T)$ at $T ≤$ 6 K evidences a slight downturn feature below $T_c^{onset} \approx$ 3 K at 5.4 GPa. Upon further increasing pressure, the drop of resistance becomes stronger and the $T_c^{onset}$ moves gradually to higher temperatures, reaching a nearly constant value of ~5 K at $P ≥$ 9.1 GPa. The zero resistance is finally achieved at $T_c^{zero} \approx$ 2-2.5 K at $P ≥$ 11.4 GPa. We found that the resistance drop in $R_a(T)$ can be suppressed by applying either high electric currents and/or external magnetic fields, as seen in Fig. 4(a) and (b). These observations are consistent with the occurrence of SC in HfTe$_3$ at $P ≥$ 5.4 GPa when the CDW transition along the $a$-axis is suppressed completely. It should be noted that the resistance data became too noisy under magnetic fields higher than 0.3 T, which prevented us from determining precisely the upper critical field.

In order to ensure the reproducibility of the experimental results, we performed HP resistance measurements on a third HfTe$_3$ (#3) single crystal along the $a$-axis. As shown in Fig. 3(d, e), the CDW transition at 93 K is clearly observed for $R_a(T)$ at 2.5 GPa and is completely suppressed at 5 GPa; the SC transition is also confirmed at low temperature for $P \geq 5$ GPa. It is noted that the observed $T_c$ for this sample is slightly lower than that of sample #2, implying that the pressure-induced SC in HfTe$_3$ is sensitive to the sample quality. As shown in Supplementary Fig. 5, the RRR values of sample #2 are indeed larger than those of sample #3 in the pressure range when CDW is suppressed completely. Accordingly, the $T_c^{onset}$ of sample #2 is also higher than that of sample #3. Nevertheless, the results of sample #3 are in general agreement with those of the #2 sample. Furthermore, we also measured the low-temperature resistance at 11 and 12.5 GPa under different electric currents and external magnetic fields. As shown in Fig. 4(c, d), the gradual suppression of the $R_a(T)$ drop upon applying high electric currents and/or external magnetic fields further substantiates the emergence of pressure-induced SC in HfTe$_3$ along the $a$-axis.

To verify if the observed SC is bulk or not, we have attempted to measure the ac magnetic susceptibility $\chi'(T)$ of HfTe$_3$ together with a piece of Pb as a superconducting reference in the CAC up to 12 GPa. As shown the Supplementary Fig. 6, no obvious diamagnetic signal was observed for HfTe$_3$ down to 1.4 K, except for the superconducting transition of Pb, which shifts to lower temperatures progressively with increasing pressure. The observed sharp transitions of Pb in Supplementary Fig. 5 elaborate an excellent hydrostatic pressure condition in the CAC at least up to 12 GPa. The absence of obvious diamagnetic response thus indicates a non-bulk or filamentary nature for the observed SC in HfTe$_3$ at least down to 1.4 K. This observation motivated us to further investigate the effect of pressure on the $b$- and $c$-axis resistances, which lead to very surprising results as shown below.

### C. Effect of pressure on the *b*-axis resistance

Because the samples #2 and #3 measured above are always disassembled into many filaments after decompression from high pressures, presumably due to the special sample microstructure and the presence of residual stress upon compression. This issue prevents the subsequent measurements along the $b$-axis direction on the identical samples. Under such a circumstance, we selected another two HfTe$_3$ samples #4 and #5 for measurements of the $b$-axis resistance $R_b(T)$ in two separated runs in the pressure range 7-12.5 GPa, where zero resistance can be achieved in $R_a(T)$. The results of $R_b(T)$ for these two samples are shown in Fig. 5(a). As seen clearly, the $R_b(T)$ curves of both samples do not exhibit any anomaly at the temperatures of ~4-5 K where $R_a(T)$ starts to show the superconducting transition. For sample #5, we observed a slight downturn of $R_b(T)$ starting at $T < 2$ K under $P \geq 10$ GPa, Fig. 5(b), which is likely caused by some misalignment of the electrical contact so that some $a$-axis component was picked up.

These results are striking. First of all, they immediately exclude the possibility that the pressure-induced resistivity drop in $R_a(T)$ arises from a secondary phase. Secondly, the observed SC in $R_a(T)$ is very likely a Q1D phenomenon. As shown below, our simulations with a 2D resistor matrix can well reproduce the above $R_b(T)$ results. To substantiate this conclusion, we further measured the $c$-axis resistance $R_c(T)$ under various pressures.

Here, we want to point out that it is quite challenging experimentally to collect merely the *b*-axis resistance without any *a*-axis contribution for the resistance measurements on a bulk sample. This is particularly difficult given the fact that the *a*-axis resistance exhibits a pronounced drop around 4-5 K whereas the *b*-axis resistance almost levels off. Therefore, a slight misalignment of the voltage leads V+/V- with respect to the *a*-axis, *i.e.*, they should be perfectly parallel with the *a*-axis, will inevitably introduce the *a*-axis feature, overshadowing the intrinsic response of *b*-axis resistivity. Under such a circumstance, the chance to observe a resistivity drop along the *b*-axis is high, but it should be convincing if we can observe no obvious resistivity drop by chance. To overcome this problem, we tried to first pattern the electrical contacts by evaporating the gold pads on the sample surface so as to make them as parallel to the *a*-axis as possible, see Supplementary Fig. 4(b), then we attached the electrical leads with silver paste. In this way, we are able to observe no any resistance drop for sample #4 and a tiny drop for sample #5 as shown in Fig. 5(b).

### D. Effect of pressure on the *c*-axis resistance

We employed the configuration shown in the inset of Fig. 5(c) to measure precisely the $R_c(T)$ of the plate-like HfTe$_3$ crystal (#6). The electric current flows through the sample via the U-shaped I+/I- pads on the opposite sides of the crystal, while the voltage leads V+/V- are attached to the center pads on both sides. At ambient pressure, $R_c(T)$ exhibits an upturn upon cooling through $T_{CDW}$ [2]. As seen in Fig. 5(c), the resistance upturn below $T_{CDW}$ is quickly suppressed by pressure, and the $R_c(T)$ at 2.4 GPa displays a metallic behavior in the whole temperature range with a clear kink-like anomaly at $T_{CDW} \approx 96.5$ K, which is consistent with that determined from $R_a(T)$ in Fig. 3(b). The CDW transition cannot be discerned any more at $P \geq 4.2$ GPa, implying the complete suppression of CDW by pressure. In the investigated pressure range, we observed no emergence of possible SC down to 1.4 K. Interestingly, the $R_c(T)$ curves at $P \geq 5.8$ GPa even exhibit an unusual upturn below around 2.5 K, as shown in Fig. 5(d). The upward trend of $R_c(T)$ below 2.5 K is more obvious when the pressure increases to 11.2 GPa. As shown below, these results can be well accounted for by turning on a Q1D SC in the resistor-matrix model. These above results thus unambiguously demonstrate the emergence of an extremely anisotropic SC in HfTe$_3$ under pressures.

### E. T-P phase diagram of HfTe$_3$

The pressure dependencies of $T_{CDW}$ and $T_c$ determined from $R_a(T)$ are shown in Fig. 6, which depicts clearly the competition between CDW and the possible Q1D SC along the *a*-axis. With increasing pressure, $T_{CDW}$ initially increases, reaching a broad maximum at 1.5 GPa, then decreases quickly and vanishes completely at about 5 GPa, above which the Q1D SC emerges abruptly. The $T_c^{onset}$ increases gradually and reaches a nearly constant value of ~ 5 K at least up to 12 GPa. The zero resistance is eventually realized at $T_c^{zero} \approx 2$ K under $P > 11$ GPa, however, the diamagnetic response cannot be obtained due to the lack of phase coherence along the *b*-and *c*-axis.

Therefore, our present work not only demonstrates unambiguously a competitive nature between CDW and SC but also reveals an intriguing anisotropic SC in the pressurized HfTe$_3$, *i.e.*, the Q1D SC emerges along the *a*-axis in a relatively broad temperature/pressure range after the CDW is completely suppressed, whereas the phase coherence between the superconducting chains cannot be realized down to 1.4 K along the other directions.

One noteworthy feature of the phase diagram shown in Fig. 6 is that SC emerges along the *a*-axis only after the CDW is suppressed completely by pressure without showing any coexistence region. In contrast, the polycrystalline HfTe$_3$ sample with at a lower $T_{CDW} \approx 80$ K shows the coexistence of CDW and filamentary SC at ambient pressure [2]. Although both the chemical disorders/defects and the physical pressure can suppress CDW and then induce SC, the above comparison actually highlights their distinct roles in tuning the CDW and SC states. In comparison with the "clean" tuning of physical pressure, the chemical disorders/defects may introduce extra charge carriers. According to the theoretical study by Imada *et al*. [23], the relationship between CDW and SC in a Q1D system depends sensitively on the band filling, *i.e.* they are separated for half-filled band but can coexist for quarter filling. They also pointed out that the coexistence of SC and CDW in a non-half-filled system is a general feature of anisotropic systems. It is plausible that the Q1D bands of HfTe$_3$ along the *a*-axis is close to half filling, leading to a complete competitive relationship between SC and CDW, while the deviation from the half-filling in the polycrystalline samples gives rise to the coexistence of CDW and SC. Our simple calculations on the density of states (DOS) distribution of the $p_x/p_y/p_z$ type Wannier orbits on the -Te2-Te3- atoms seem to support such a scenario, *i.e.*, the filling numbers of the $p_x$-type Wannier orbits along the 1D atom chain direction are about 0.55, Supplementary Fig. 14. It should be noted that the valence state of Te ions in HfTe$_3$ is complex, and more experimental and theoretical investigations are needed to understand the interplay between CDW and SC.

### F. Crystal structure under high pressure

To study the structural stability of HfTe$_3$ under compression, we also measured the high-pressure XRD on the pulverized HfTe$_3$ crystals at room temperature and found no structure transition up to at least 14 GPa. As shown in Fig. 7, the lattice parameters *a*, *b*, *c*, and volume *V* contract smoothly with increasing pressure, implying that the crystal structure is stable in the studied pressure range. The pressure dependence of volume *V*(*P*) can be well fitted with the Birch-Murnaghan (B-M) equation of state as shown by the solid line in Fig. 7(b). By fixing $B_1 = 4$, we obtained a bulk modulus $B_0 = 45.3(7)$ GPa and $V_0 = 228.7(4)$ Å$^3$. In addition, there is no obvious anomaly for the lattice parameter *a* at $P_c \approx 5$ GPa, where the CDW disappears and the Q1D SC emerges at low temperatures. The axial ratios *c/a* and *b/a* exhibit the monotonic decrease with the pressure gradually increasing (see Supplementary Fig. 7), indicating the *b* and *c* axis is more compressible than *a*-axis. The smooth evolutions of the axial ratios *c/a* and *b/a* up to ~ 10 GPa also indicate that the crystal structure is stable within the studied pressure range. Therefore, we can safely conclude that pressure-induced SC around 5 GPa should be correlated with an electronic transition rather than a structural transition,

*i.e.*, it is linked directly with the elimination of the CDW order. In this sense, the pressure-induced SC along the *a*-axis can be regarded as a new electronic phase, which shares similar crystal structure as the ambient phase but has no CDW instability.

**Discussions**

In order to simulate the above anisotropic transport behaviours at low temperatures under HP, we designed a simplified 2D resistor matrix consisting of 45 resistors as shown in Supplementary Figs. 8-11. We first simulate the change of *b*-axis resistance $R_b$ with a perfect alignment upon entering the superconducting state along the *a*-axis, manifested by a reduction of the resistors' resistance from 5 to 0.1 Ω along this direction. As shown in Supplementary Fig. 8, the voltage drop across V+/V- along the *b*-axis (or equivalently $R_b$) remains unchanged when the resistance of resistors along the *a*-axis are reduced from 5 to 0.1 Ω, consistent with the experimental results for sample #4 in Fig. 5(b). However, when the alignment of voltage leads is not perfect, some *a*-axis component will be picked up as simulated in Supplementary Fig. 9. In this case, the voltage drop across V+/V- is found to decrease slightly when the resistors along the *a*-axis are reduced from 5 to 0.1 Ω. This reproduces well the observed resistance drop of sample #5 shown in Fig. 5(b), confirming that the change of $R_a$ will affect the $R_b$ sensitively if the alignment of voltage leads is imperfect.

We further simulated the influence of resistance drop of $R_a$ and the misalignment of V+/V- on the *c*-axis resistance $R_c$. As illustrated in Supplementary Figs. 10 and 11, interestingly, the voltage drop across V+/V- along the *c*-axis (or equivalently $R_c$), no matter aligned perfectly or not, was both found to rise progressively when the resistance of the resistors along the *a*-axis is reduced gradually, in excellent agreement with the experimental results shown in Fig. 5(d). In contrast, the voltage drop will decrease if the superconducting transition takes place in the *c*-axis direction. These above simulations not only reproduce the experimental observations, but also support strongly the occurrence of a Q1D SC in the pressurized HfTe3 single crystal.

To further substantiate the occurence of Q1D SC along the *a*-axis, we then analyzed the excess condcutivity $(\Delta\sigma = \sigma - \sigma_0)$ around $T_c$ according to A-L model [21], *viz*.

$$\Delta\sigma \propto [(T - T_c)/T_c]^{-(4-D)/2} \quad (1)$$

where *D* is the dimension and $T_c$ is the mean-field transition temperature, which equal to $T_c^{zero}$. For *D* =1, the excess conductivity due to 1D superconducting fluctuations is given by:

$$\Delta\sigma = (\pi e^2/16\hbar d^2)\,\xi(0)[(T - T_c)/T_c]^{-3/2} \quad (2)$$

Here, *e* is the electron charge, *d* is the diameter of a superconducting filament, and $\xi(0)$ is the coherence length at *T* = 0 K. As illustrated in Fig. 8, the excess conductivity $\Delta\sigma$ from 9 to 12.5 GPa for sample #3 measured at a smaller current 0.1 mA indeed follows a nice linear relationship as a function of $[(T - T_c)/T_c]^{-3/2}$; here the $\Delta\sigma$ is calculated from data shown in Fig. 3(e) and $\sigma_0$ is the conductivity at 6 K for *I*//*a* with

0.1 mA current. These observations indicate that strong 1D superconducting fluctuations persist even up to 12.5 GPa. It is also noteworthy that the slope of the fitting curve decreases gradually with increasing pressure with a ratio of $k_{9\text{ GPa}}:k_{11\text{ GPa}}:k_{12.5\text{ GPa}} = 2.4:1.4:1$. As shown in Fig. 3(e), the $T_c$ in the investigated pressure range only changes slightly, and thus the change of $\xi(0)$ should not dominate the significant decrease of the slope, or the pre-factor of the above Eq. (2). Instead, the reduction of the slope should be mainly attributed to the increase of the diameter $d$ of the superconducting filaments under pressure. This is conceivable considering the fact that the superconducting filaments are getting closer under HP. It can be expected that the superconducting state will gradually evolve into three dimensions under higher pressure or lower temperatures where the Josephson coupling between filaments are strengthened.

Before discussing the possible origins for the observed Q1D SC in the pressurized HfTe$_3$, it is impertive to understand the nature of the CDW transition and its interplay with SC. A side-by-side comparison between HfTe$_3$ and ZrTe$_3$ is also instructive. At ambient pressure, ZrTe$_3$ and HfTe$_3$ exhibit a CDW transition at $T_{\text{CDW}}$ = 63 K and 93 K, respectively. The determiend CDW vector of HfTe$_3$, i.e. $\boldsymbol{q}$ = 0.91(1) $a$* + 0.27(1) $c$* is very similar to that of ZrTe$_3$, i.e. $\boldsymbol{q}$ = 0.93 $a$*+ 0.33 $c$* [6]. In general opinion, the CDW vector is related with both the FS nesting vector and the electron-phonon coupling distribution. As shown in Fig. 9, the calculated electronic susceptibility $\chi_{\boldsymbol{q}}'$ of HfTe$_3$ demonstrates an obvious peak at 0.90 $a$*, which is in excellent agreement with the experimental result determined from the TEM measurements. Nevertheless, there is no obvious peak along the $c$* direction, so it is difficult to determine the $c$* component of $\boldsymbol{q}$ merely based on the information of FS nesting. We suspect that the electron-phonon coupling in HfTe$_3$ may play a dominant role in the $c$* component of its CDW vector. In comparison, the calculated electronic susceptibilities of ZrTe$_3$ under 0 and 5 GPa are shown in Supplementary Fig. 13. The red regions were drawn by the contour line of calculated electronic susceptibility $\chi_{\boldsymbol{q}}'$, while the experimentally determined CDW vectors $\boldsymbol{q}$ were also labeled. Obviously, the $c$* components of experimental $\boldsymbol{q}$ do not correspond to the peaks of calculated $\chi_{\boldsymbol{q}}'$, while their $a$* components seem to be well matched. This is similar to the above case of HfTe$_3$.

Below $T_{\text{CDW}}$, an anisotropic superconducting transition was discovered in ZrTe$_3$ [5,13]. In specific, $\rho_a$ starts to decrease at $T_c^{\text{onset}} \approx$ 4 K and reaches zero at $T_c^{\text{zero}} \approx$ 2 K with a relatively broad superconducting transition, while $\rho_b$ exhibits a much narrower transition from $T_c^{\text{onset}} \approx$ 2 K to $T_c^{\text{zero}} \approx$ 1.5 K. Based on the specific-heat measurements, the observed anisotropic superconducting transition has been ascribed to a gradual crossover from a filamentary SC (between 4 and 2 K) induced by local pairs along the $a$-axis to a bulk SC (below 1.5 K) when the superconducting filaments form phase coherence [13]. Such a mixed bulk-filament SC observed in ZrTe$_3$ has been attributed to the peculiar electronic structures characterized by the presence of Q1D and 3D FSs [10,13,19]. In striking contrast to ZrTe$_3$, no SC was detected down to 50 mK for the high-quality HfTe$_3$ single crystals [2]. Their distinct low-temperature behaviors can be rationalized by comparing to the effects of physical pressure on the CDW and SC of ZrTe$_3$. At ambient conditions, the cell volume of HfTe$_3$ (228.70 Å$^3$) is about 1.35% smaller than that of ZrTe$_3$ (231.83 Å$^3$),[8,20] which corresponds to the application of 0.84 GPa hydrostatic pressure on ZrTe$_3$ given a compressibility of 0.016 GPa$^{-1}$ [14]. At this pressure, it was found that the CDW transition of ZrTe$_3$ is enhanced to ca. 100 K while the SC is suppressed completely [19,22], which are similar to the situations seen in HfTe$_3$ at ambient pressure.

These above simple arguments indicate that the coexistent CDW and SC in ZrTe$_3$ at ambient pressure are actually competing with each other, which becomes more evident under pressure. For ZrTe$_3$, a reentrant SC emerges above 5 GPa when the CDW disappears abruptly; the observed SC was believed to be bulk in nature [17], but the diamagnetic evidence is still lacking to date as mentioned above [12]. In the present case, we observed similar pressure-induced SC in $R_a$ of HfTe$_3$, but the absences of SC along the other directations and the absence of diamagnetic signals point to an intrinsic filamentary character for the observed SC. Our work thus calls for a reexamination of the anisotropic properties of ZrTe$_3$ under high pressures. Our DFT calculations also provide some clues for the suppression of CDW under pressure. As shown in Fig. 9(c, d), the distances between the Te2 and Te3 atoms in the -Te2-Te3- chain along the *a*-axis becomes more uniform at high pressures. In addition, there are more localized electrons between the Te2 and Te3 atoms from different prismatic (HTe$_3$)$_\infty$ chains.

Finally, we discuss the possible origins for the observed exteremly anistorpic or Q1D SC in the pressurized HfTe$_3$ according to the proposal by Yamaya *et al.*[13] Similar to ZrTe$_3$ [8,9], our electronic structure calculations show that HfTe$_3$ owns both 3D FSs around the BZ center and Q1D FSs around the BZ boundary, as shown in Fig. 10(a). The former ones are dominated by Hf 5*d* orbitals in the (HfTe$_3$)$_\infty$ prism, while the latter ones are mainly contributed by unidirectional Te-5$p_x$ orbitals within the -Te2-Te3-Te2-Te3- chain along the *a*-axis, Fig. 1(a). At the intersection area between these two types of FSs, there are also Q1D+3D FSs as indicated by the black arrows in Fig. 10(a). On these FSs, the Fermi velocities are displayed with the color scales, where the red and blue colors represent the highest and zero Fermi velocities, respectively. In particular, the low Fermi velocities near the Q1D+3D FSs with narrow band feature may result in van Hove singularity (vHs) in the density of states. It is known that the presence of a vHs at $E_F$ can induce FS instabilities and lead to SC or CDW states. In the narrow bands that construct the Q1D+3D FSs, electrons can form local pairs that interact attractively with each other over short distances [13]. With increasing pressure, as shown in Fig. 10(b), the 3D FSs of HfTe$_3$ around the Brillouin zone center expand, which enhance their overlapping with the Q1D FSs and reduce the composition of the Q1D bands at the Fermi level. As a result, the FS nesting of Q1D FSs along the *a\** direction is weakened under pressure, which may suppress the CDW order and induce the SC. Since the narrow-band electron of HfTe$_3$ also has a dominant Te-5$p_x$ characteristic originating in the -Te2-Te3- chains as in ZrTe$_3$, the local pairs must be dominantly formed along the *a*-axis, leading to the Q1D SC transition in the *a*-axis direction [19]. In comparison, the 3D FSs of ZrTe$_3$ are larger than those of HfTe$_3$ at 0 GPa and even become the open-orbital pockets along the $a^*$ direction at 5 GPa, as shown in Fig. 10 (c-d). This indicates that the CDW order due to the Q1D FSs nesting in ZrTe$_3$ is not so strong as in HfTe$_3$, which is consistent with the tendency of the experimental $T_{CDW}$ of ZrTe$_3$ (63 K) [6] versus HfTe$_3$ (93 K). Therefore, the anistropic SC property in HfTe$_3$ can maintain in a wider temperature/pressure range, which makes HfTe$_3$ being an ideal platform to study the local pairs in the Q1D superconducting system.

In summary, we have measured the anisotropic resistances of the CDW conductor HfTe$_3$ under high pressures. We found that a Q1D SC emerges in $R_a(T)$ at $T < 4\sim5$ K when the CDW is suppressed completely by $P > 5$ GPa, whereas no clear sign of SC is observed down to 1.4 K in $R_b(T)$ and $R_c(T)$. Our results demonstrate an intimate interplay between CDW and SC in HfTe$_3$, which are particularly interesting given their Q1D characters. We have compared our results with those of ZrTe$_3$ and discussed the pressure-induced Q1D SC in pressurized HfTe$_3$ in terms of the peculiar electronic structures and the local pairs formed along the -Te2-Te3- chains parallel to the $a$-axis.

## Methods

### Single-crystal growth and characterization

High-quality HfTe$_3$ single crystals used in the present study were grown with the iodine chemical-vapor-transport method as described elsewhere [2]. Before HP measurements, the crystals have been carefully characterized at ambient pressure by single-crystal X-ray diffraction (XRD), energy dispersive spectroscopy (EDS), TEM, and resistivity measurements. The CDW vector $q$ at ambient pressure was determined by TEM measurements at 26 K. The specimens for TEM observation were prepared by mechanical exfoliation at room temperature and then transfer onto the Cu grid with the aid of crystalbond and acetone. In-situ low-temperature TEM experiments were performed using a liquid-helium-cooled specimen holder (GATAN, HCHDT3010) in a JEM-2100F electron microscope equipped with a charge-coupled device (CCD) camera system and operated at 200 keV.

### Details of high-pressure measurements

The temperature dependences of resistance $R(T)$ under various pressures were measured with the standard four-probe method by using either a self-clamped piston-cylinder cell (PCC) [24] in the pressure range $0 < P \leq 2$ GPa or a palm-type cubic anvil cell (CAC) [25] in the pressure range $2 < P \leq 12$ GPa. In order to improve the electrical contact, we first evaporated four gold pads on the sample surface in vacuum and then attached four gold wires of 20 μm in diameter on top of the gold pads with the silver paste. AC magnetic susceptibility under pressure has been measured with the mutual induction method. The excitation and pick-up coils are made of enameled cooper wires of 25 μm in diameter, and they were manually wound around a Teflon sheet with a rectangular cross section of 0.3 mm × 0.5 mm. The Teflon sheet can be removed easily and the rectangular space ensures a large filling factor for the susceptibility measurements. Each coil has two layers and about 40 turns in total. The HfTe$_3$ sample and a piece of lead (Pb) with similar size of about 0.3 × 0.3 × 0.25 mm$^3$ were put inside the pick-up coil. An excitation current of 1 mA with frequency of 1117 Hz was applied to the excitation coil and the output signal as picked up with a Stanford Research SR830 lock-in amplifier. For these HP measurements, the Daphne 7373 or glycerol was employed as the pressure transmitting medium. The pressure in PCC was determined from the superconducting transition temperature of Pb, while the pressure in CAC was estimated from the calibration curve at low temperatures [26]. High-pressure synchrotron powder XRD ($\lambda$ = 0.6199 Å) was performed at room temperature at 4W2 beamline, Beijing Synchrotron Radiation Facility (BSRF). Silicon oil was used as the transmitting medium. The Le Bail method was employed to fit the XRD data with the General Structure Analysis System (GSAS) program package program. [27] The ruby fluorescence method was used to determine the pressure.

**Theoretical calculations**

To investigate the electronic structures of HfTe$_3$ under pressure, we carried out density functional theory (DFT) calculations with the projector augmented wave method [28, 29] as implemented in the VASP package [30-32]. The generalized gradient approximation of Perdew-Burke-Ernzerhof (PBE) type [33] was adopted for the exchange-correlation functional. The kinetic energy cutoff of the plane wave basis was set to 300 eV. A 16 × 16 × 8 $k$-point mesh was used for the Brillouin zone (BZ) sampling. The Fermi surface was broadened by the Gaussian smearing method with a width of 0.05 eV. The spin-orbit coupling effect was included as both Hf and Te are heavy elements. The vdW interaction was described by the optB86b-vdW functional [34]. In the structural optimization, both cell parameters and internal atomic positions were allowed to relax until the forces on all atoms were smaller than 0.01 eV/Å. The Fermi surface (FS) and Fermi velocity of HfTe$_3$ were studied with the Wannier90 [35] and FermiSurfer [36] packages, respectively. To describe the Fermi surface nesting, the real part of electronic susceptibility $\chi_{\boldsymbol{q}}'$ ($\chi_{\boldsymbol{q}}' = \sum_{k,ij} \frac{f(\varepsilon_{k+q,j}) - f(\varepsilon_{k,i})}{\varepsilon_{k,i} - \varepsilon_{k+q,j}}$) was computed, where $f(\varepsilon_{k,i})$ is the Fermi-Dirac distribution function and $\varepsilon_{k,i}$ is the energy of band $i$ at vector $\boldsymbol{k}$.

**Data Availability**

The datasets generated during and/or analyzed during the current study are available from the corresponding author on reasonable request.

**Acknowledgments**

We are grateful to the enlightening discussions with Prof. Jianshi Zhou and Dr. Jiaqiang Yan. This work is supported by the Beijing Natural Science Foundation (Z190008), National Natural Science Foundation of China (12025408, 11921004, 11888101, 11834016, 11904391, 11874400, 11774424), the National Key R&D Program of China (2018YFA0305702, 2018YFA0305800, and 2017YFA0302903), the Strategic Priority

Research Program and Key Research Program of Frontier Sciences of the Chinese Academy of Sciences (XDB25000000, XDB33000000 and QYZDB-SSW-SLH013), and the CAS interdisciplinary Innovation Team (JCTD-2019-01) as well as the Users with Excellence Program of Hefei Science Center CAS (Grant No. 2021HSC-UE008). Y.U. acknowledges the support from JSPS KAKENHI (19H00648). ADXRD measurements were performed at 4W2 High Pressure Station, Beijing Synchrotron Radiation Facility (BSRF), which is supported by Chinese Academy of Sciences (Grant KJCX2-SW-N20, KJCX2-SW-N03).

## Author contributions

J.G.C. directed this research. J.L., S.Z., and G.F.C. grew the $HfTe_3$ single crystals. Z.Y.L. and P.T.Y. performed the high-pressure measurements and data analyses. J.L. and H.X.Y. carried out the TEM measurements and data analysis. Y.U. provided the high-pressure methodology. J.F.Z. and K.L. did the electronic structure calculations. Z.Y.L., K.L., Y.S. and J.G.C wrote this manuscript with comments from all authors.

## Competing interests

The authors declare no competing financial or non-financial interests.

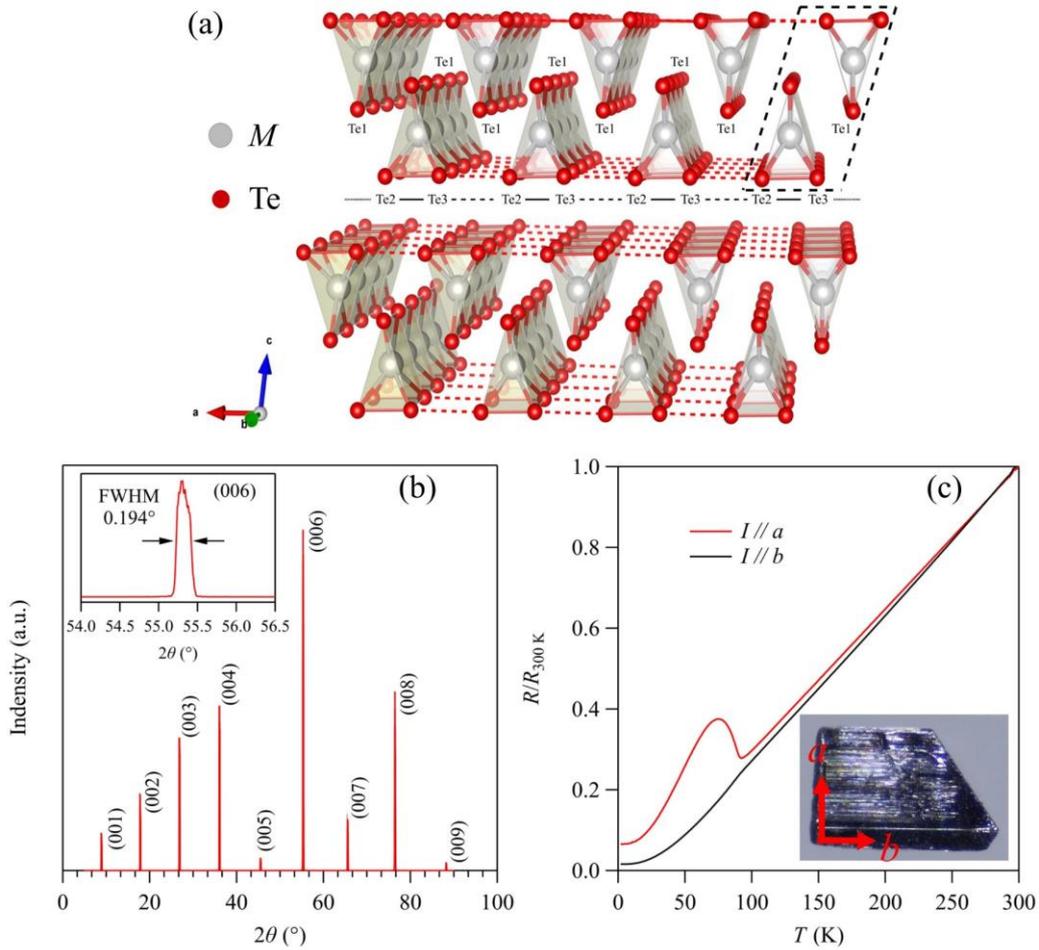

Fig. 1 (a) Crystal structure of $M$Te$_3$ ($M$ = Zr, Hf); (b) X-ray diffraction pattern of the single-crystal HfTe$_3$ showing only the (00$l$) diffraction peaks. The inset shows the FWHM of the (0 0 6) Bragg peak; (c) Temperature dependence of the $a$- and $b$-aixs normalized resitance $R(T)/R_{300\ \text{K}}$ measured on the same piece of sample at ambient pressure. Displayed in the inset of (c) is the photo of the crystal, which shows fringes extending along the $b$-axis.

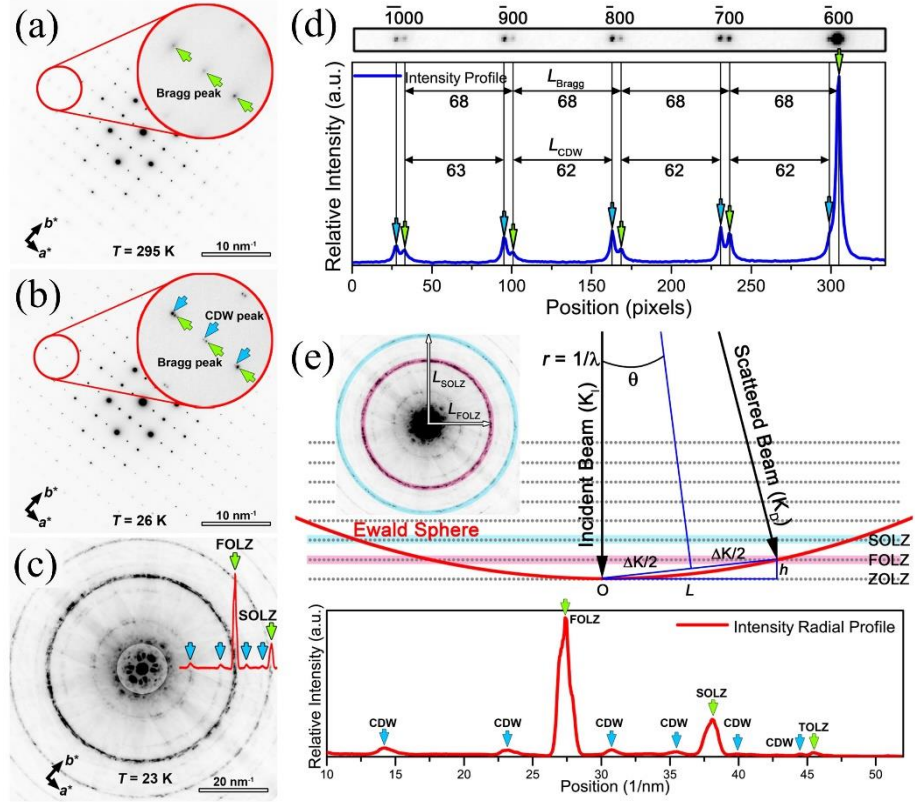

Fig. 2 Electron diffraction patterns showing the superlattice in HfTe$_3$. Typical selected area electron diffraction (SAED) patterns taken along the [001] zone axis at (a) 295 K and (b) 26 K, respectively. The satellite spots (indicated by blue arrows) appear at low temperatures around the Bragg spots (green arrows), indicating the formation of structural modulation. (c) A convergent beam electron diffraction (CBED) pattern taken along the [001] zone axis, showing extra Laue zones (blue arrows) due to the structural modulation. The radial intensity distribution is plotted as an inset, with the positions of the first/second order Laue zone (FOLZ/SOLZ) indicated by green arrows. Based on (b) and (c), the modulation vector can be determined to be $q = 0.91(1)\ a^* + 0.27(1)\ c^*$. (d) Intensity line profile from spot ($\bar{6}$00) to ($\overline{10}$00) extracted from the SAED pattern (b). The intervals between different spots are labeled. (e) Upper panel: the geometric configuration showing the Ewald sphere intercepts points in the Laue zones. Lower panel: the intensity radial profile extracted from CBED pattern (c), from which the $L$ values can be obtained.

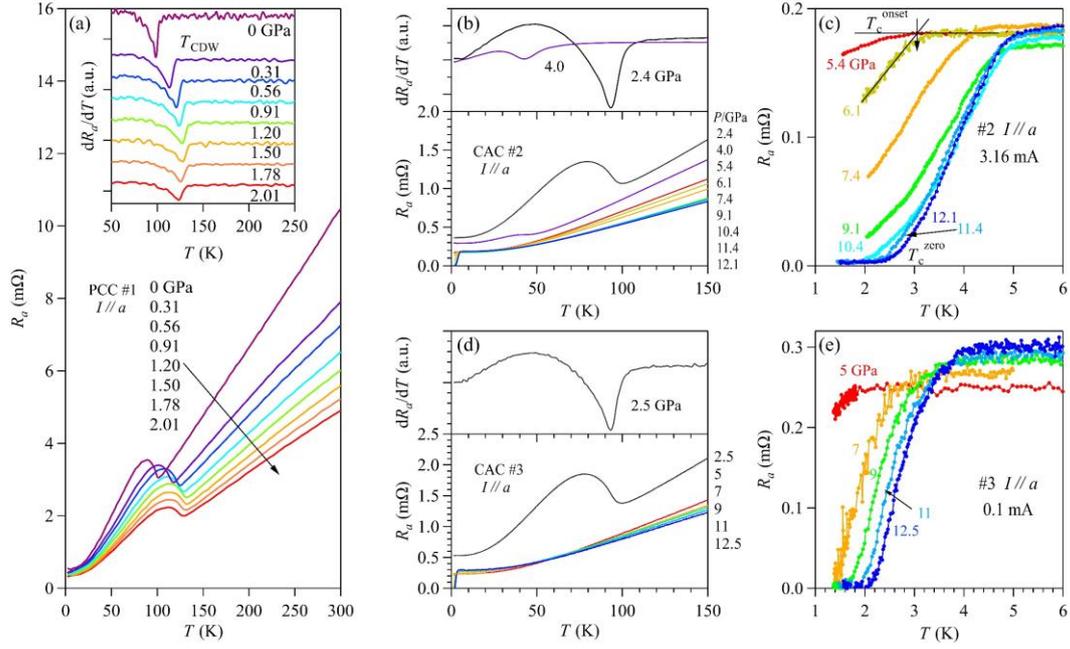

Fig. 3 (a) Temperature dependence of the $a$-axis resistance $R_a(T)$ for sample #1 under various pressures up to 2.1 GPa measured with a piston-cylinder cell (PCC); the inset shows the $dR_a/dT$ of the data in (a) to determine the charge-density-wave transition temperature $T_{CDW}$. (b) $R_a(T)$ for sample #2 under various pressures from 2.4 to 12.1 GPa measured with a palm cubic anvil cell (CAC); the upper panel shows the $dR_a/dT$ curves at 2.4 and 4.0 GPa to determine the values of $T_{CDW}$; (c) The low-temperature $R_a(T)$ for sample #2 under various pressure from 5.4 to 12.1 GPa. (d) $R_a(T)$ for sample #3 under various pressures from 2.5 to 12.5 GPa; the upper panel shows the $dR_a/dT$ curve at 2.5 GPa to determine the values of $T_{CDW}$; (e) The low-temperature $R_a(T)$ for sample #3 under various pressure from 5 to 12.5 GPa highlighting the evolution of the possible superconducting transition.

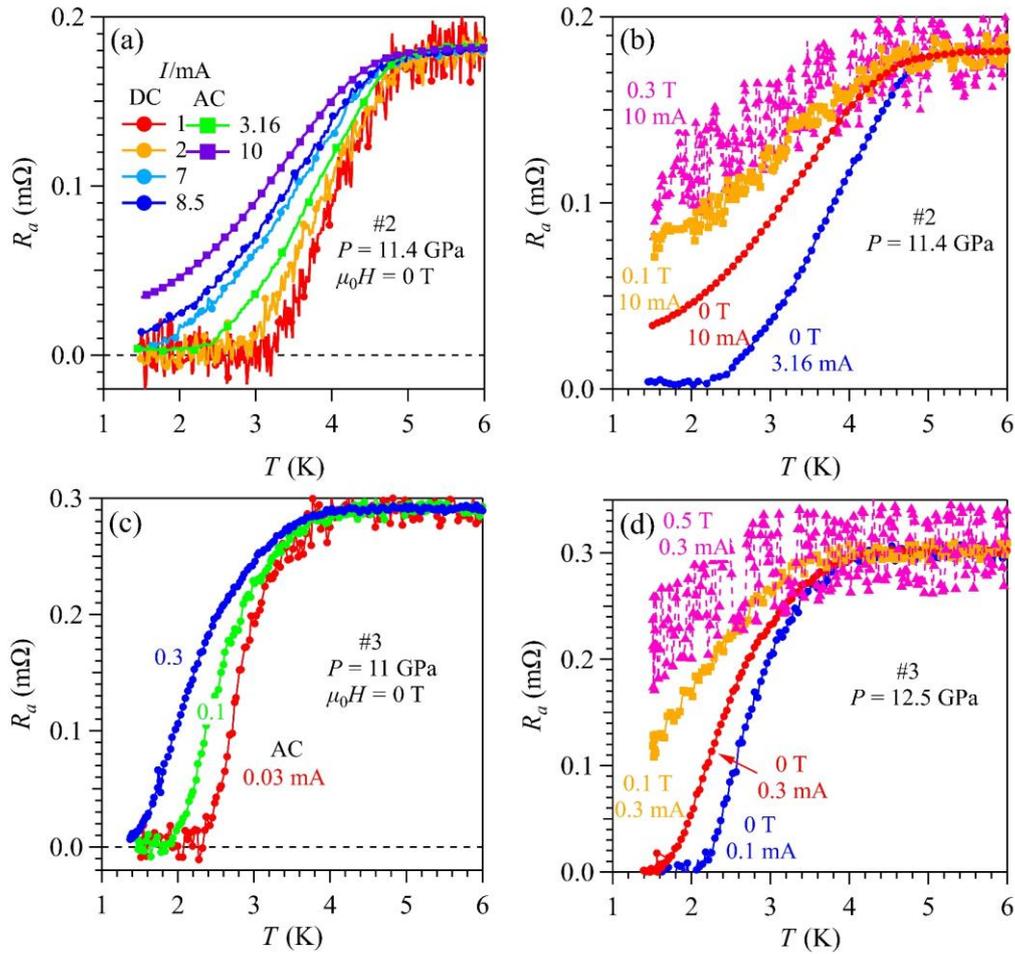

Fig. 4 The low-temperature $R_a(T)$ for sample #2 at 11.4 GPa measured with (a) different currents and (b) magnetic fields; (c) The low-temperature $R_a(T)$ for #3 at 11 GPa measured with different currents; (d) The low-temperature $R_a(T)$ for #3 at 12.5 GPa measured with different currents and/or magetic fields.

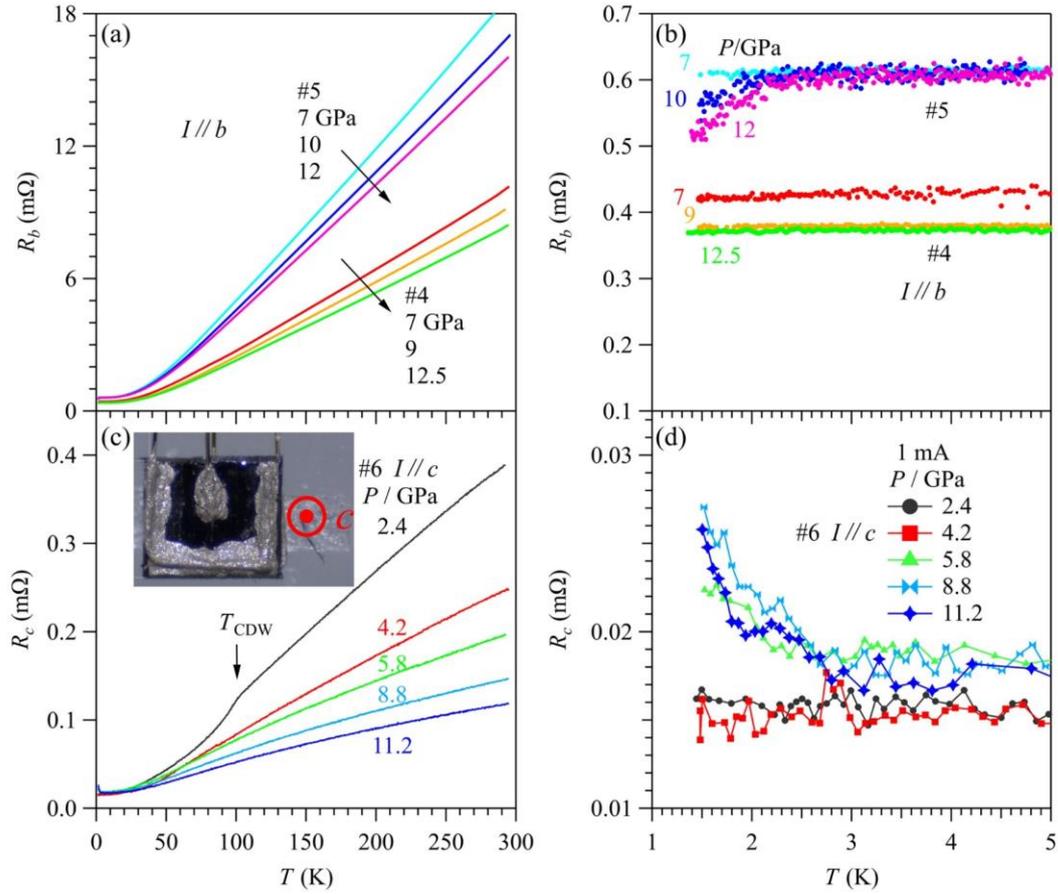

Fig. 5 (a) The temperature dependence of $b$-axis resistance $R_b(T)$ under different pressures from 7 to 12.5 GPa for two samples (#4 and #5) and (b) their low-temperature data. (c) The temperature dependence of $c$-axis resistance $R_c(T)$ under different pressures from 2.5 to 11 GPa for sample #6 and (d) their low temperatrue parts below 5 K. The inset is a schematic diagram of the resistance mearuesment along $c$-aixs setup.

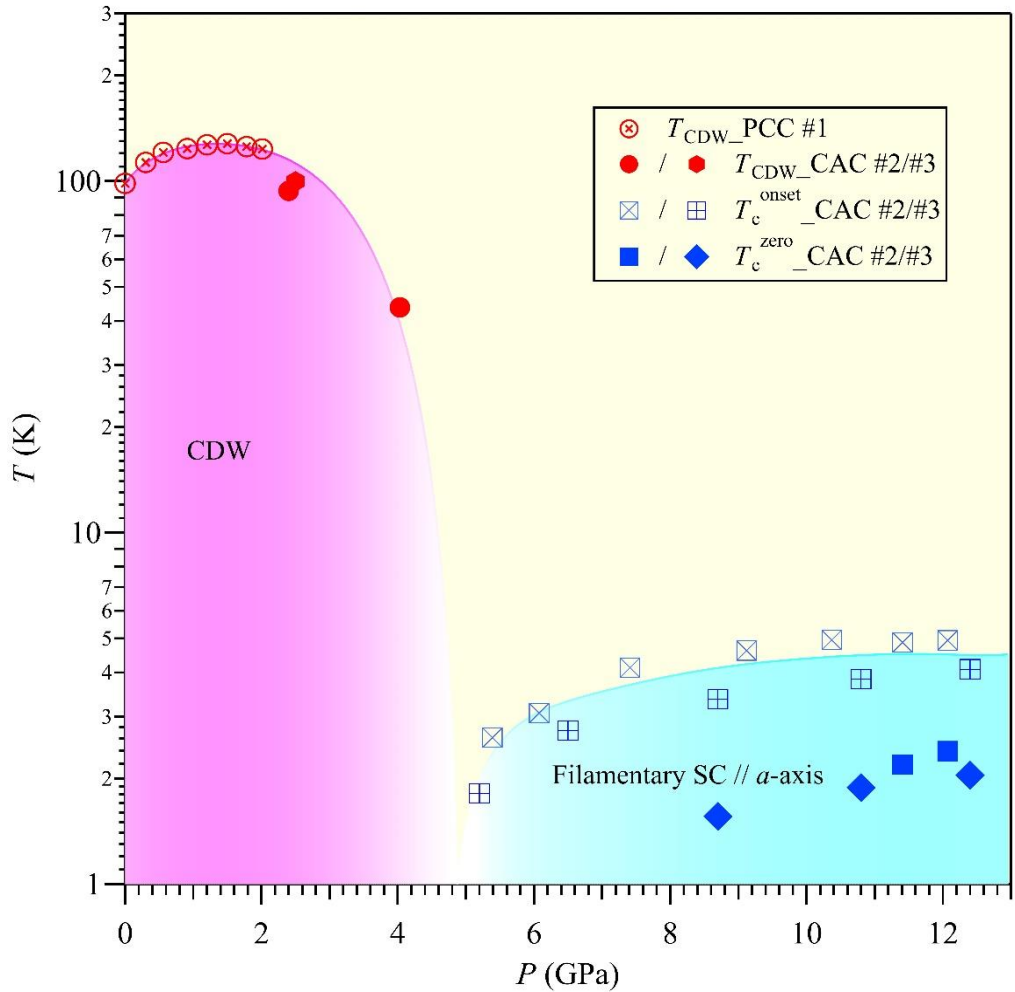

Fig. 6 Temperature-pressure phase diagram of HfTe$_3$ single crystal.

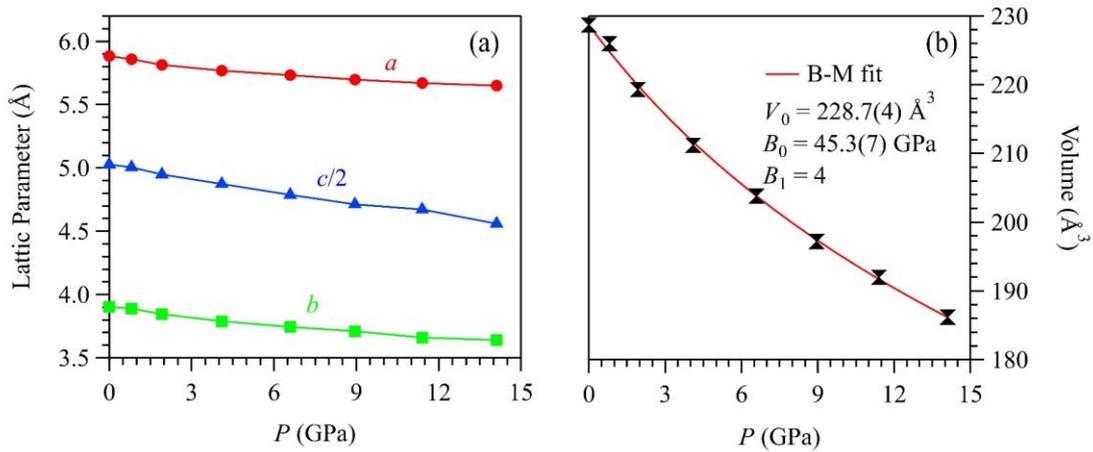

Fig. 7 Pressure dependences of (a) the lattice parameters and (b) volume of HfTe$_3$ at room temperature. The solid line in (b) represents the fit with the Birch-Murnaghan (B-M) equation of state to extract the bulk modulus $B_0$ as given in the figure.

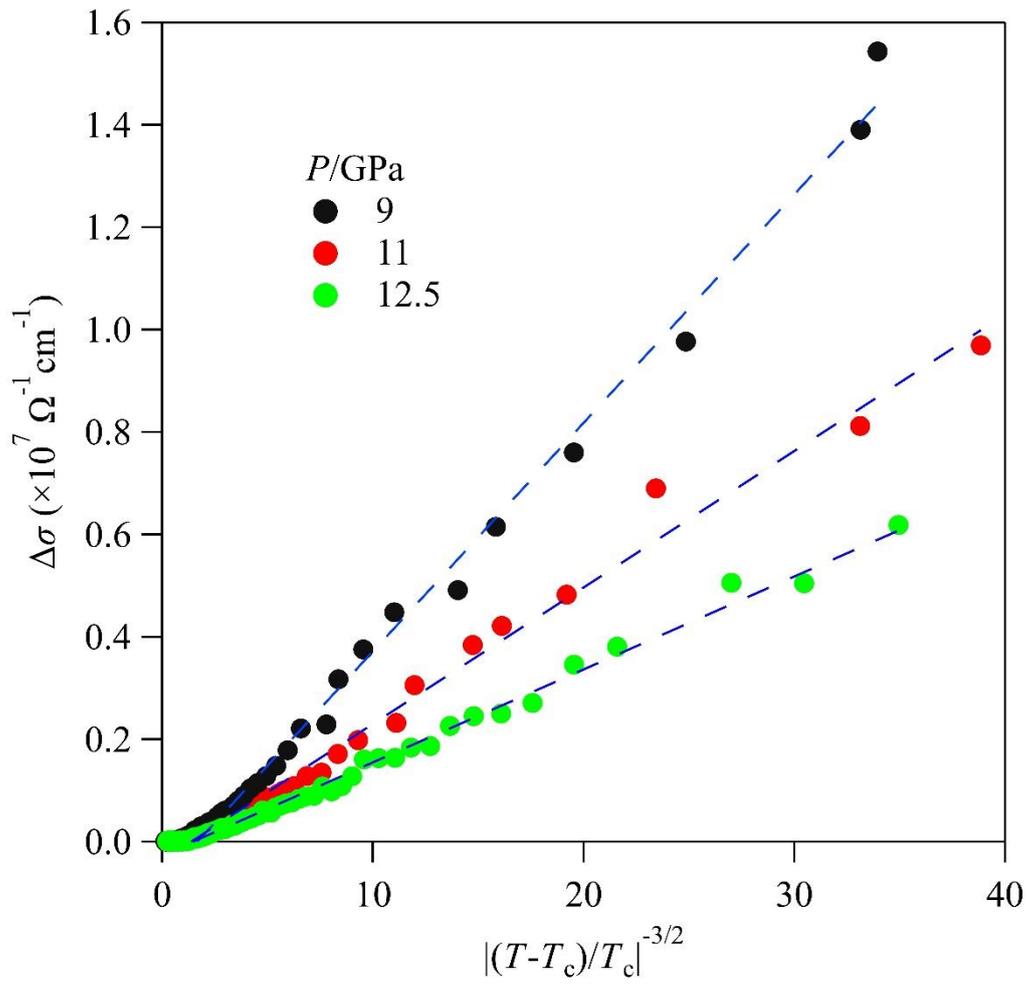

Fig. 8 The reduced temperature dependence of excess conductivity around $T_c$ along the *a*-axis.

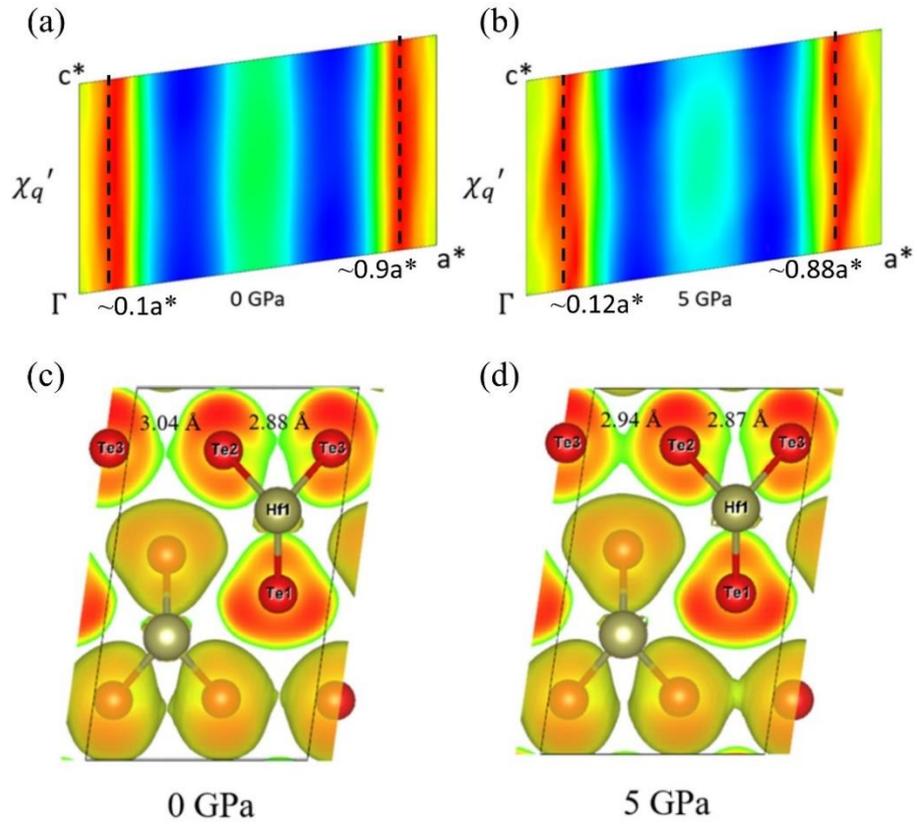

Fig. 9 The calculated real part of the electronic susceptibility $\chi_q'$ for HfTe$_3$ in the $a^*$-$c^*$ plane at (a) 0 GPa and (b) 5 GPa. The components along the $b^*$ direction have been integrated. The calculated electronic localization function of HfTe$_3$ at (c) 0 GPa and (d) 5 GPa. The isosurface was set as 0.5.

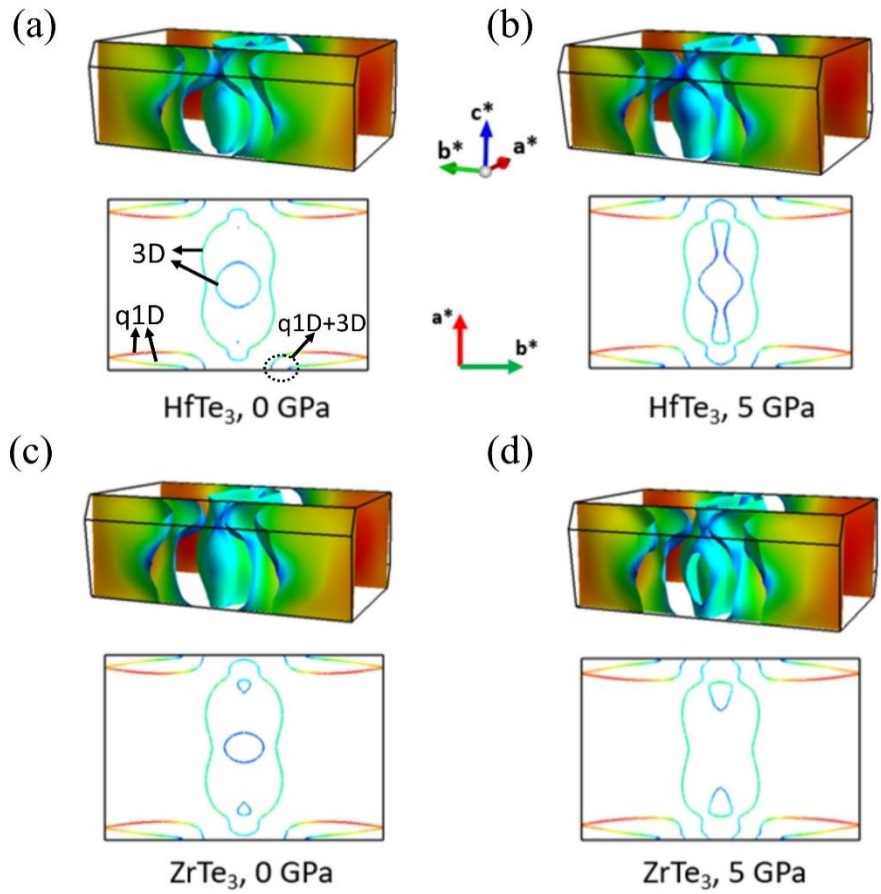

Fig. 10 The calculated Fermi surface sheets of (a-b) HfTe$_3$ and (c-d) ZrTe$_3$ at 0 GPa and 5 GPa, respectively. The color scales on the Fermi surface represent the Fermi velocities, where the pure red and pure blue correspond to the highest and zero Fermi velocities, respectively. The bottom figures in each panel are the sections of Fermi surface sheets at the $c^* = 0$ plane.

# Supplementary Materials

# Quasi-one-dimensional superconductivity in the pressurized charge-density-wave conductor HfTe$_3$

Z. Y. Liu[1,2#], J. Li[2,3#], J. F. Zhang[4#], J. Li[2,3], P. T. Yang[2], S. Zhang[2,3*], G. F. Chen[2,3], Y. Uwatoko[5], H. X. Yang,[2,3] Y. Sui[1*], K. Liu[4*], and J.-G. Cheng[2,3*]

[1]*School of Physics, Harbin Institute of Technology, Harbin 150001, China*

[2]*Beijing National Laboratory for Condensed Matter Physics and Institute of Physics, Chinese Academy of Sciences, Beijing 100190, China*

[3]*School of Physical Sciences, University of Chinese Academy of Sciences, Beijing 100190, China*

[4]*Department of Physics, Beijing Key Laboratory of Opto-electronic Functional Materials & Micro-nano Devices, Renmin University of China, Beijing 100872, China*

[5]*Institute for Solid State Physics, University of Tokyo, Kashiwa, Chiba 277-8581, Japan*

\# These authors contributed equally to this work.

*E-mails: jgcheng@iphy.ac.cn; szhang@iphy.ac.cn; suiyu@hit.edu.cn; kliu@ruc.edu.cn

## Supplementary Notes1: Single crystal characterizations

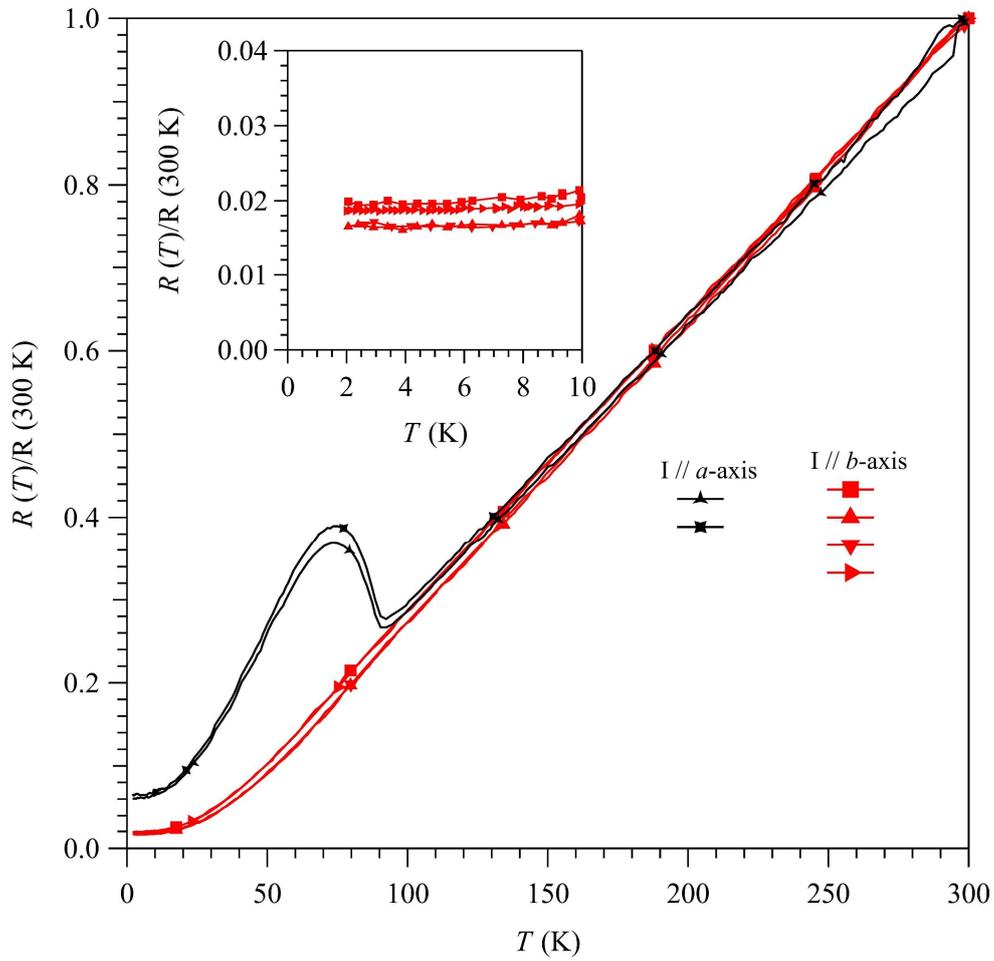

**Supplementary Figure 1.** Normalized resistance of several selected HfTe$_3$ samples from room-temperature down to 2 K measured along *a*- and *b*-axis. The inset shows an enlarged view of the low-temperature region along the *b*-axis. The residual resistivity ratio ($R_{300K}/R_{2K}$) along the *b*-axis for all samples are larger than 50, indicating a high quality of the selected samples for high-pressure transport measurements. This is further elaborated by the absence of resistivity drop down to 2 K as shown in the inset.

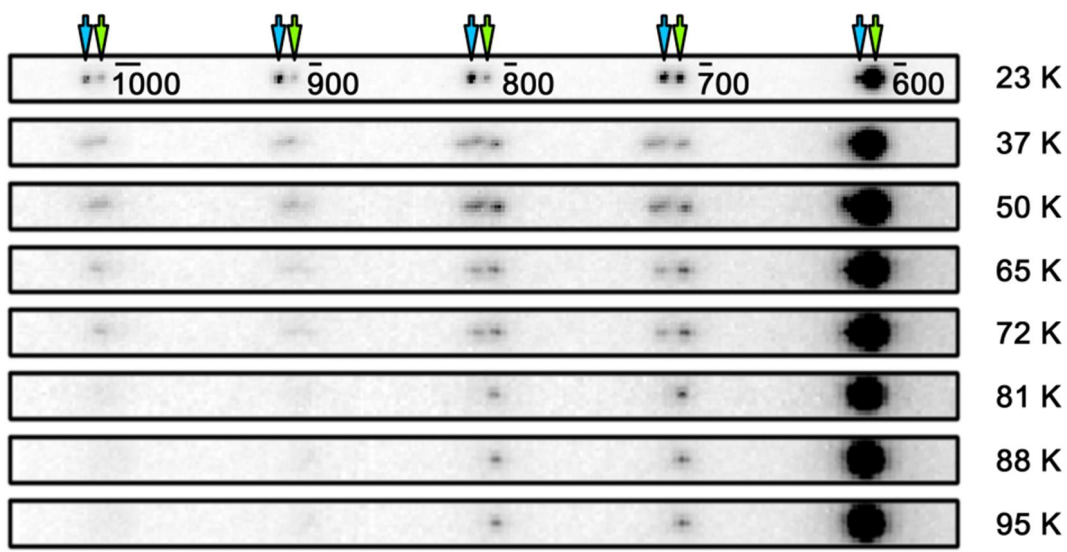

**Supplementary Figure 2.** Temperature dependence of electron diffraction patterns of HfTe$_3$ across the CDW phase transition. The Bragg (from ($\bar{6}$00) to ($\bar{10}$00)) and CDW spots are indicated by green and blue arrows, respectively.

The stronger intensities of the CDW spots compared to the adjacent Bragg spots shown in **Figs. 2(b, d)** of the main text can be interpreted with the aid of the following Supplementary Fig. 3. The relaxation of the Bragg conditions occurs at the reciprocal lattice points in the case of thin TEM specimen. Every point in the reciprocal lattice should be replaced by a relrod elongated normal to the thin film, and the intensity recorded by the camera depends on the distance of the relrod center to the intersection with the Ewald sphere. For HfTe$_3$, $q_{CDW}$ = 0.91(1) $a^*$ + 0.27(1) $c^*$, the CDW relrods locate between the ZOLZ and FOLZ, as schematically shown in the Supplementary Fig. 3. The Ewald sphere can only intersect the CDW relrod centers at relatively large scattering angles, which is accompanied by the dimmed Bragg spots. As a result, the intensity decreases monotonously from Bragg spot ($\bar{6}$00) to ($\bar{10}$00), while the CDW intensity reaches its maximum in the vicinity of $(\bar{7}00)$, as shown in **Fig. 2(d)** of the main text.

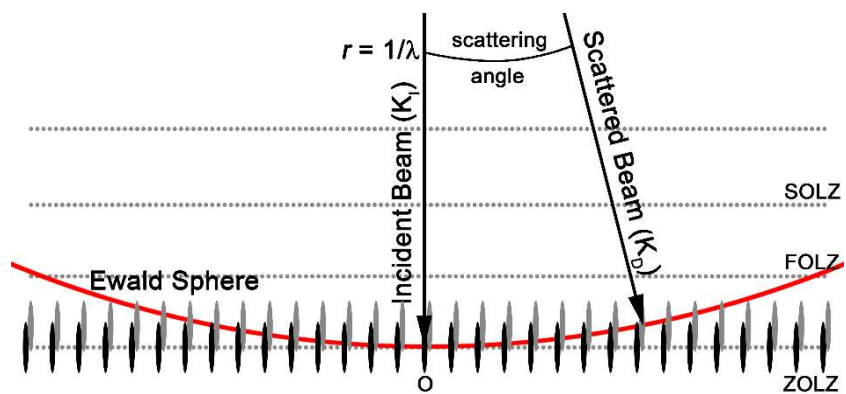

**Supplementary Figure 3.** Schematic of the Ewald sphere and diffraction conditions. The intensity recorded depends on the distance of the relrod center to the intersection with the Ewald sphere. The black and gray ellipses represent the Bragg and CDW relrods, respectively.

## Supplementary Notes2: High-pressure measurements

The picture of a single crystal $HfTe_3$ shown in Supplementary Fig. 4. shows fringes extending along the *b*-axis, which is parallel to the crystal edge. Then, the *a*-axis is perpendicular to the crystal edge. For our resistivity measurements along the *a*- or *b*-axis, we have cut the sample in such a way that the longest dimension of about 0.5 mm is along the axis to be measured. To improve the electrical contact, we first evaporated four parallel gold pads on the top surface of the crystal, as indicated by the red dotted frames in Supplementary Fig. 4(b), and then attached 20 μm gold wires on the pads with silver paste. Supplementary Figs. 4(b, c) show the photos of the samples measured along the *b*- and *a*-axis, respectively. The distances marked on the photos were measured directly by a stereomicroscope. It is noted that the sample along the *a*-axis, Supplementary Fig. 4(c), was cut a bit wider because a narrower sample would disassemble into many fibers due to the peculiar microstructure of the sample. We have employed the configuration shown in Supplementary Fig. 4(d) to measure the resistivity along the *c*-axis. The U-shaped I+/I- pads on the opposite sides of the plate-like crystal can ensure a uniform current flow through the crystal so that the central V+/V- pads can pick up the voltage drop along the *c*-axis. Supplementary Fig. 4(e) shows the photo of the sample measured along the c-axis.

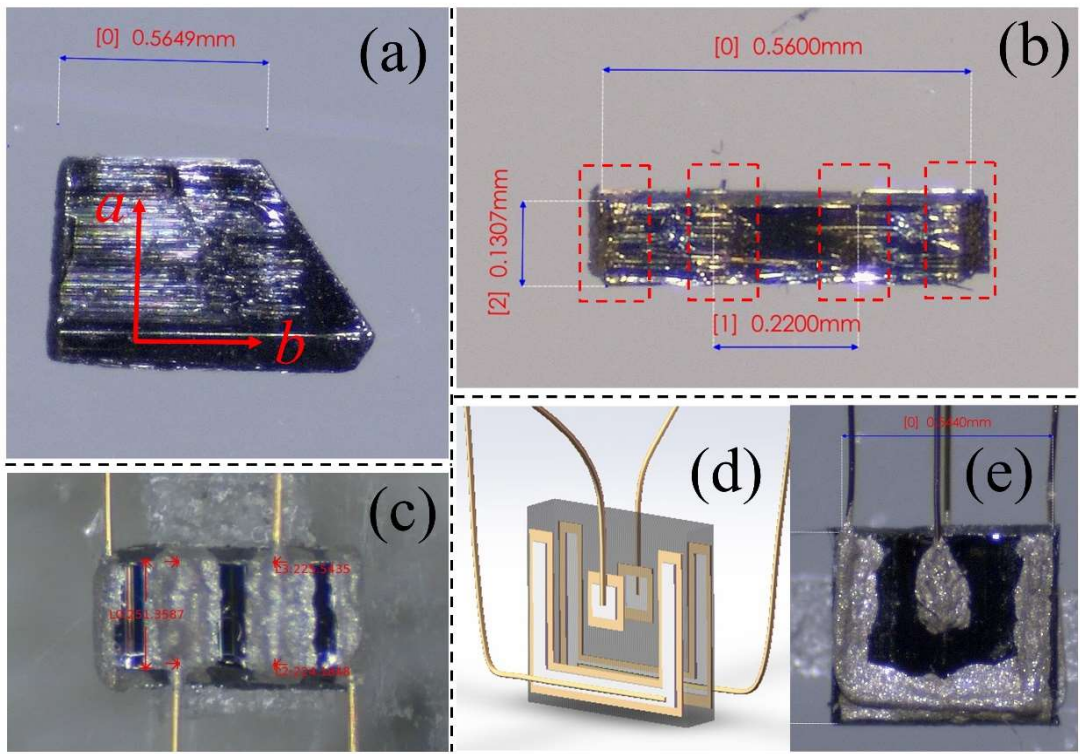

**Supplementary Figure 4.** (a) The morphology of the single-crystals HfTe$_3$ showing

clearly the fringes along the *b*-axis with the orthogonal *a*-axis also marked, (b) The *b*-axis sample showing the parallel gold pads evaporated on the surface of sample, (c) The *a*-axis sample with the gold wires attached to the gold pads, (d) and (e) are the schematic drawing and real photo of the *c*-axis sample, respectively.

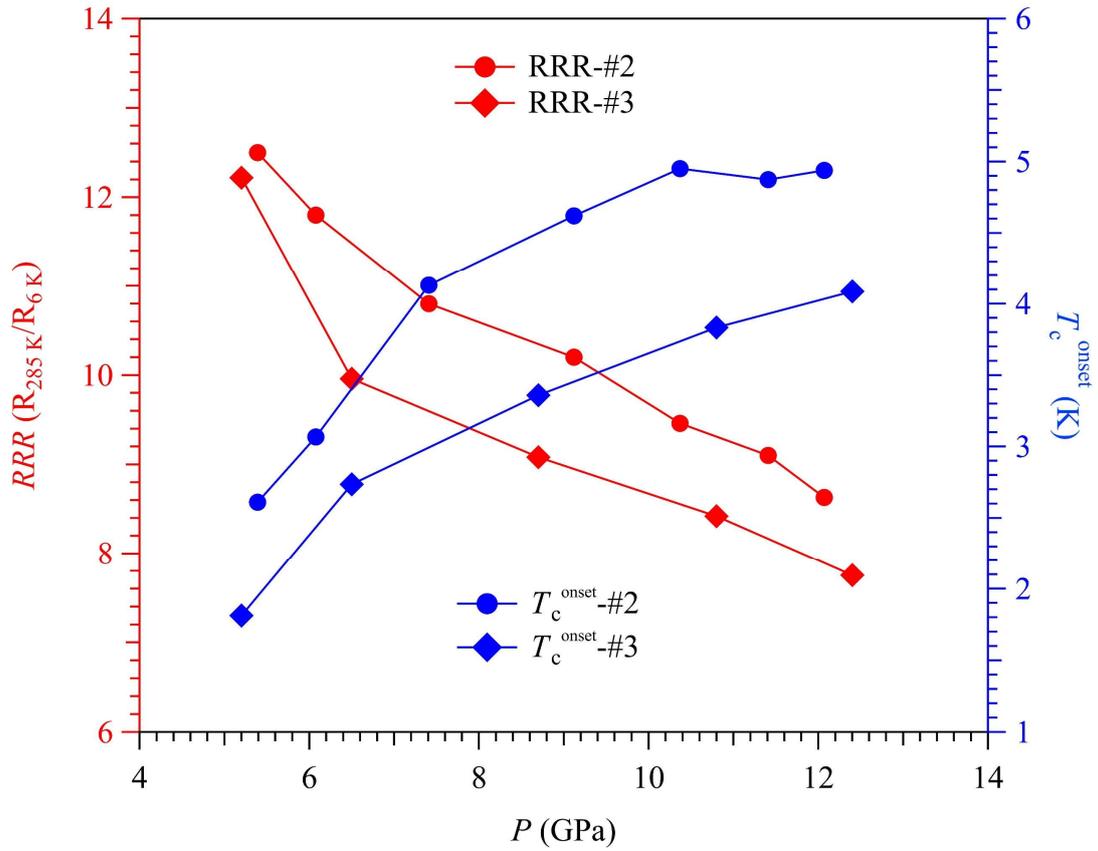

**Supplementary Figure 5.** The superconducting transition temperature $T_c^{onset}$ and residual resistivity ratio (RRR) of samples #2 and #3. The RRR of sample #2 is larger than that of sample #3. Correspondingly, the $T_c^{onset}$ of sample #2 is also higher than that of sample #3.

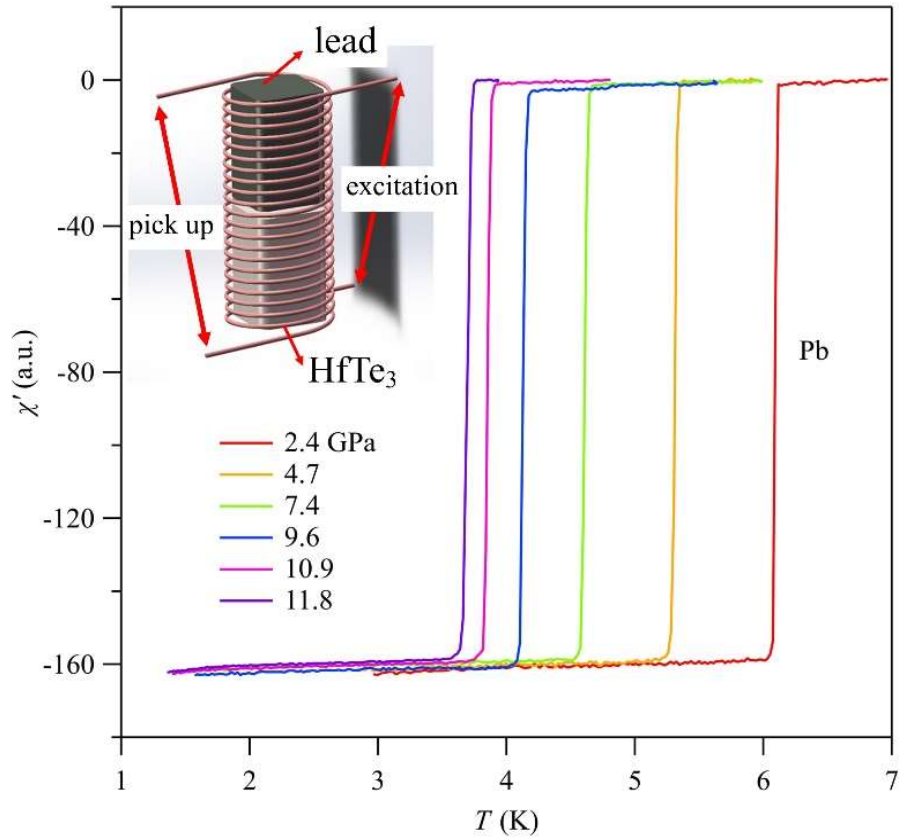

**Supplementary Figure 6.** Temperature dependence of the AC magnetic susceptibility $\chi'(T)$ of HfTe$_3$ and Pb and under various pressure up to 11.8 GPa. The inset is a schematic diagram of the AC magnetic susceptibility measurement setup. The superconducting transition of Pb moves down progressively with increasing pressure. The observed sharp transitions of Pb elaborate an excellent hydrostatic pressure condition in the CAC at least up to 12 GPa. But no obvious diamagnetic signal was observed for HfTe$_3$ down to 1.4 K, the lowest temperature in the present study. The absence of obvious diamagnetic response thus indicates a filamentary nature for the observed SC in HfTe$_3$ at least down to 1.4 K.

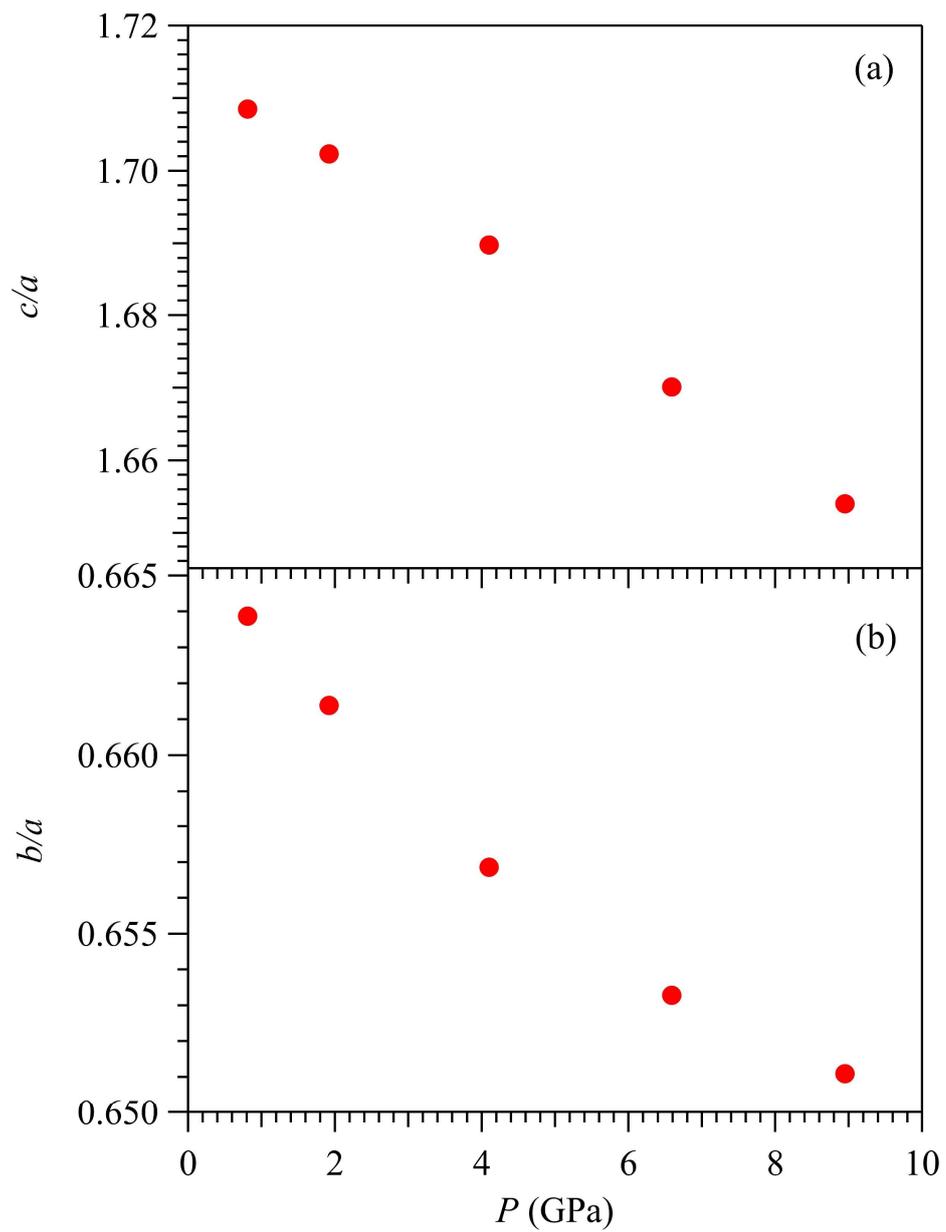

**Supplementary Figure 7.** Pressure dependences of the axial ratios *c*/*a* and *b*/*a*.

# Supplementary Notes3: Resistor-matrix simulations

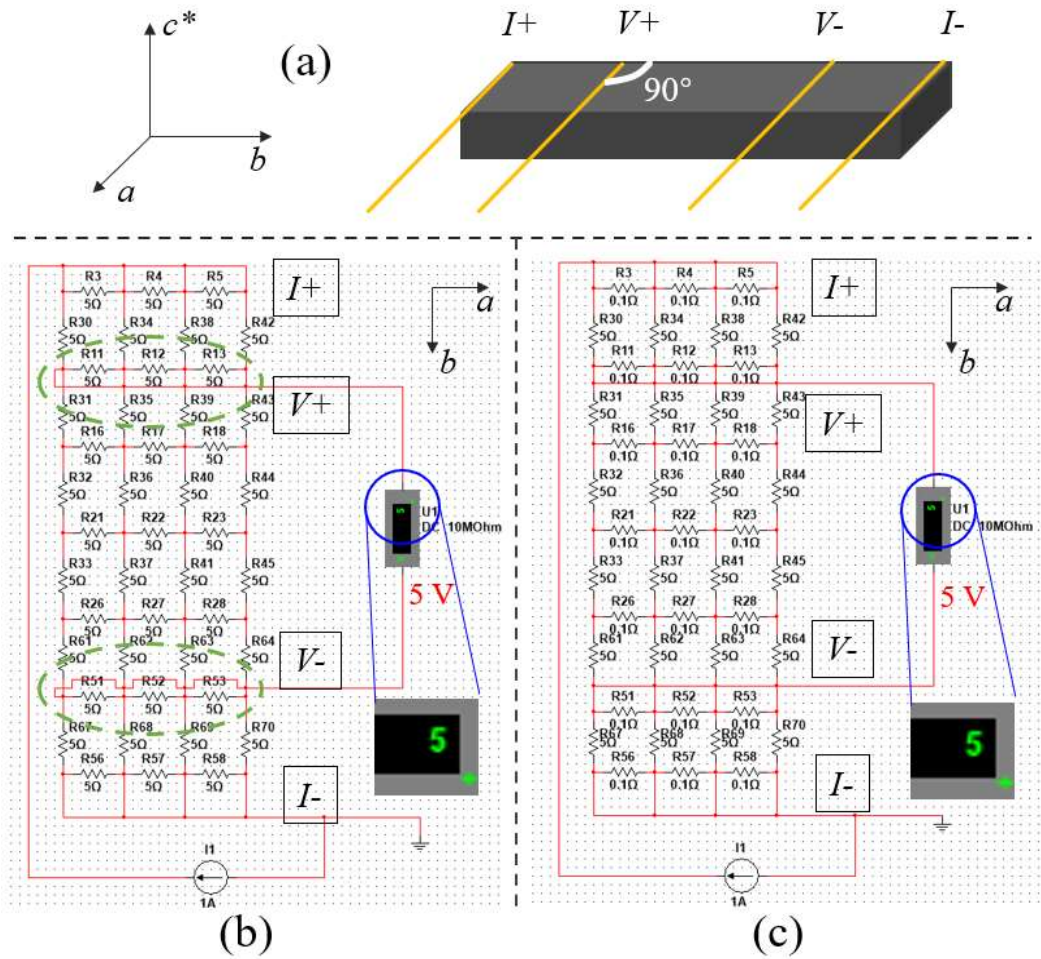

**Supplementary Figure 8.** (a) The schematic diagram of the resistance measurement along the *b*-axis. The resistor matrix and the simulation result for (b) the normal state and (c) the superconducting state along *a*-axis. Here, we apply a current of 1 A, and set every resistor as 5 Ω in the normal state, which gives a voltage drop across V+/V- of 5V as shown in (b). When the *a*-axis resistors are transformed to a superconducting state, simulated by a reduction of resistor from 5 to 0.1 Ω, the voltage drop remains at 5V as shown in (c). This result indicates that the *b*-axis resistance $R_b$ is not affected by the change of *a*-axis resistance if the voltage leads are aligned perfectly perpendicular to the *b*-axis, consistent with the experimental results for sample #4 in Fig. 5(b) of the main text.

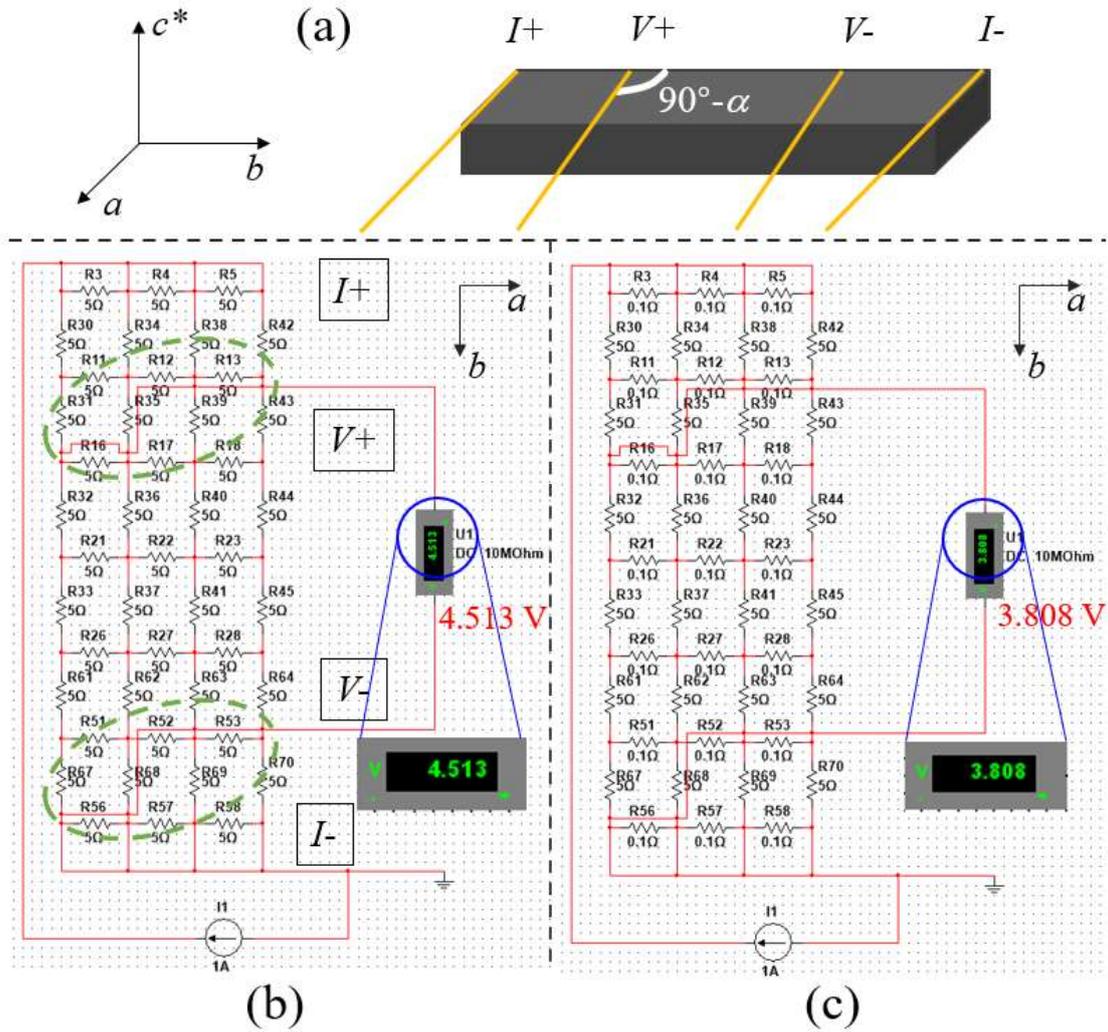

**Supplementary Figure 9.** (a) The schematic diagram of the *b*-axis resistance with imperfect alignment. The resistor matrix and the simulation result for (b) the normal state and (c) the superconducting state along *a*-axis. In this case, the imperfect alignment is simulated by passing the voltage leads across some *a*-axis resistors. As shown in (b, c), the voltage drop across V+/V- is found to decrease from 4.513 to 3.808 when the superconducting transition takes place along the *a*-axis. This result demonstrates that the change of the *a*-axis resistance will affect the *b*-axis resistance if the voltage leads are not aligned perfectly, consistent with the experimental results for sample #5 in Fig. 5(b) of the main text.

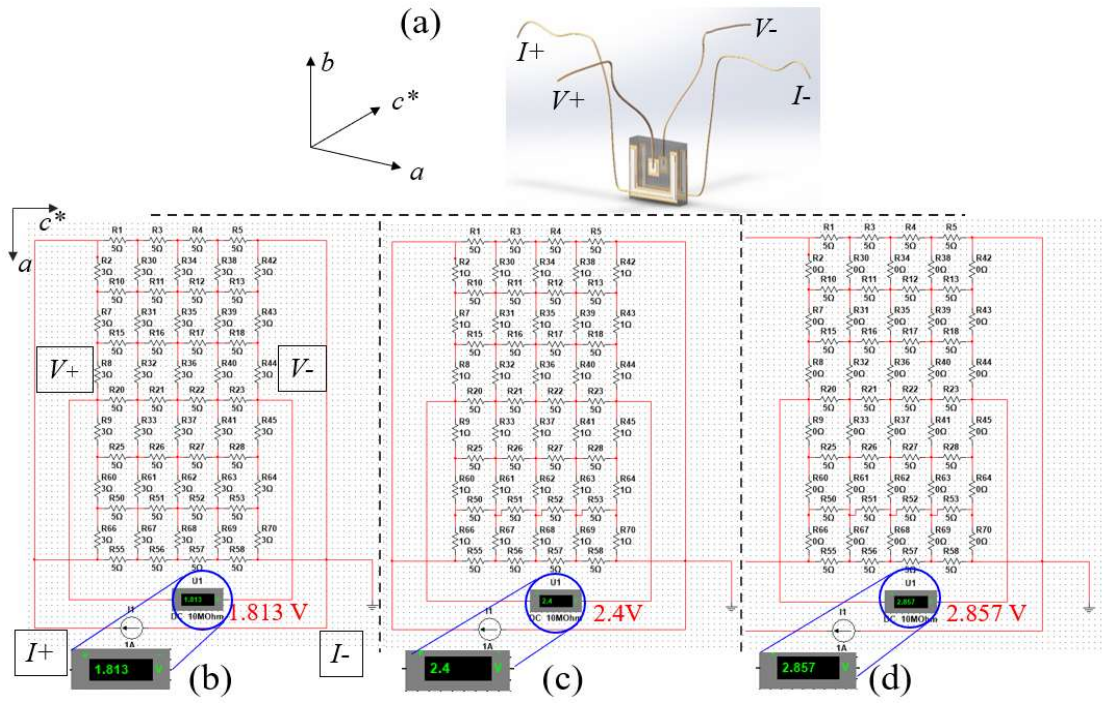

**Supplementary Figure 10.** (a) The schematic diagram of the resistance measurement along *c*-axis. (b-d) The resistor matrix and the simulation result for the *c*-axis resistance upon transforming the *a*-axis resistors from the normal state to the superconducting state.

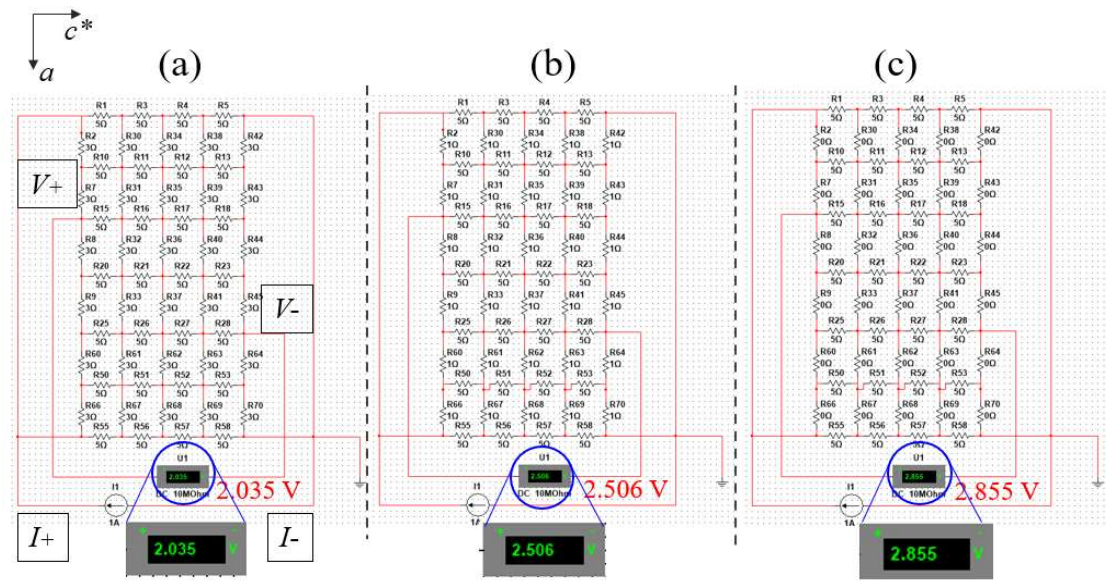

**Supplementary Figure 11.** The resistor matrix and the simulation result for the *c*-axis resistance upon transforming the *a*-axis resistors from the normal state to the superconducting state when the V+/V- are off centered.

As illustrated in Supplementary Figs. 10 and 11, the voltage drop across V+/V- along the *c*-axis (or equivalently $R_c$), no matter aligned perfectly or not, was both found to rise progressively when the resistance of the resistors along the *a*-axis is reduced gradually, in excellent agreement with the experimental results shown in Fig. 5(d) of the main text.

**Supplementary Notes4: Computational results**

In the structural optimization, both cell parameters and internal atomic positions were allowed to relax until the forces on all atoms were smaller than 0.01 eV/Å. The calculated structure parameters agree well with the experimental measurements, Supplementary Fig. 12.

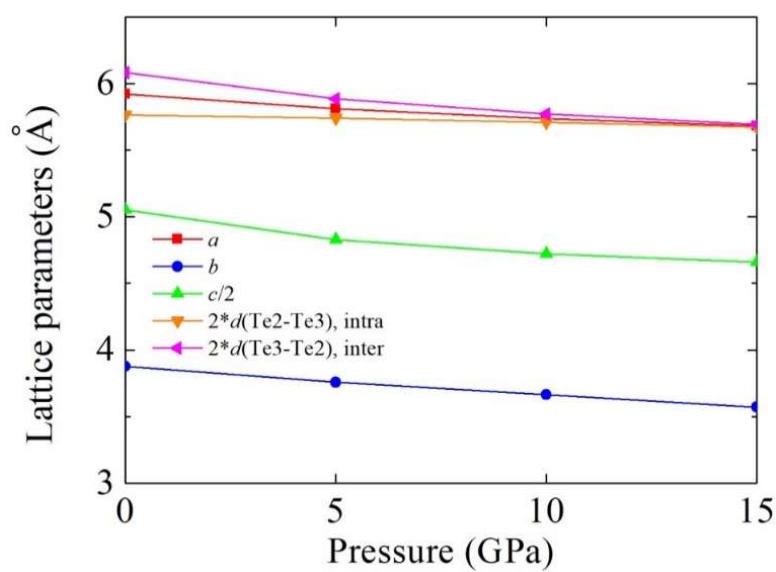

**Supplementary Figure 12.** The calculated lattice parameters of HfTe$_3$ under different pressures. Here *a*, *b*, and *c* are lattice constants, while *d* is the distance between different Te atoms.

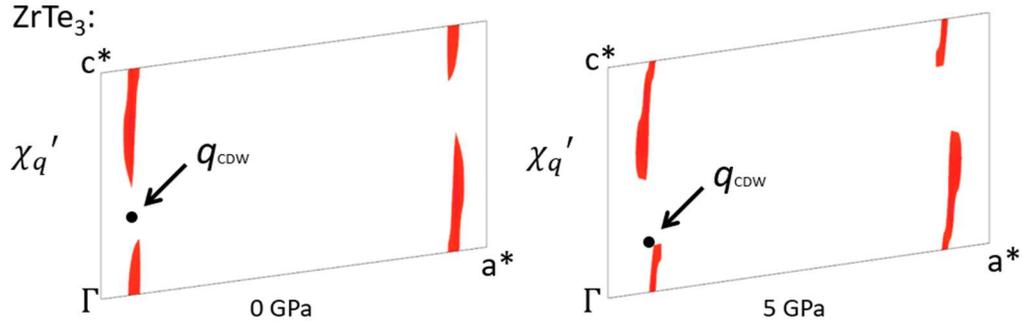

**Supplementary Figure 13.** The calculated real part of the electronic susceptibility $\chi_q'$ for ZrTe$_3$ in the $a^*$-$c^*$ plane at (a) 0 GPa and (b) 5 GPa. The components along the $b^*$ direction have been integrated. The black dots label the experimentally determined $q_{CDW}$ from Ref. [1].

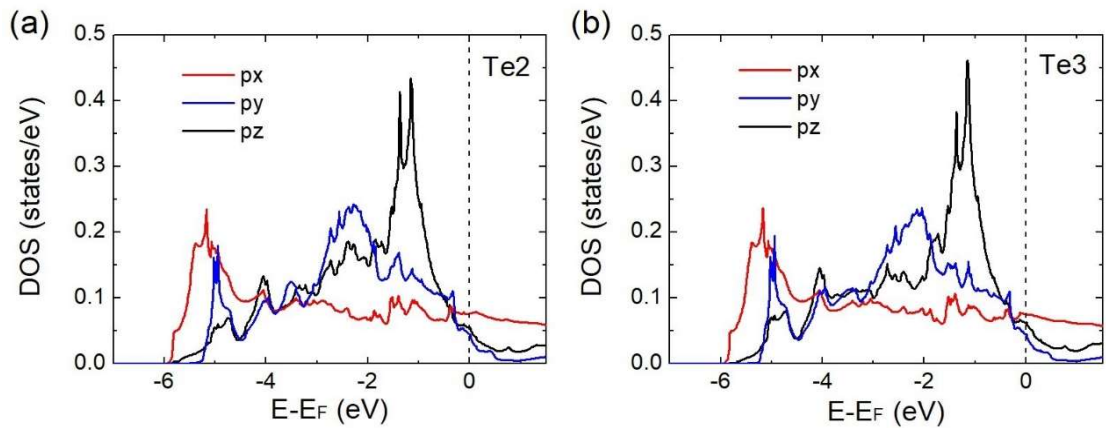

**Supplementary Figure 14.** The density of states (DOS) distribution of the $p_x$/$p_y$/$p_z$ type Wannier orbits of the Te2/Te3 atoms in the 1D atomic chain. Among them, the filling numbers of the $p_x$-type Wannier orbits along the 1D atom chain direction ($x$-direction) are about 0.55, which is approximate to the half filling case and can bring the completely competitive relationship between the SC and CDW

**Supplementary References:**